\documentclass[10pt,final,journal,twocolumn]{IEEEtran}

\ifCLASSINFOpdf

\else

\fi

\usepackage{hyperref}

\usepackage{cite}
\usepackage[cmex10]{amsmath}
\usepackage{amsmath}
\usepackage{algorithm}
\usepackage{algorithmic}
\usepackage{stfloats}
\usepackage{multicol,multienum}
\usepackage{multirow}
\usepackage[dvips]{graphicx}
\usepackage{wrapfig}
\usepackage{color}

\usepackage{cite}
\usepackage[cmex10]{amsmath}
\usepackage{amsmath}
\usepackage{algorithm}
\usepackage{algorithmic}
\usepackage{stfloats}
\usepackage{multicol,multienum}
\usepackage{multirow}
\usepackage[dvips]{graphicx}

\usepackage[tight,footnotesize]{subfigure}
\usepackage{wrapfig}
\usepackage{color}

 \usepackage{booktabs}
 \usepackage{multirow}

\usepackage{mathtools}

\usepackage{float}
\usepackage{bm}

\usepackage{pifont}

\usepackage{bbding}
\usepackage{bbding}
\usepackage{pifont}
\usepackage{wasysym}
\usepackage{amssymb}
\usepackage{amsfonts}

\begin{document}

\title{\huge Safeguarding ISAC Performance in Low-Altitude Wireless Networks Under Channel Access Attack}

    \author{
    Jiacheng Wang, Jialing He, Geng Sun, Zehui Xiong, Dusit Niyato,~\IEEEmembership{Fellow,~IEEE,} Shiwen Mao,~\IEEEmembership{Fellow,~IEEE,} Dong In Kim,~\IEEEmembership{Life Fellow,~IEEE,} Tao Xiang,~\IEEEmembership{Senior Member,~IEEE}
    \vspace{-1cm}
    
    \thanks{J.~Wang and D. Niyato are with the College of Computing and Data Science, Nanyang Technological University, Singapore (e-mail: jiacheng.wang@ntu.edu.sg, dniyato@ntu.edu.sg).}

    \thanks{J.~He and T. Xiang are with College of Computer Science, Chongqing University, Chongqing 400044, China (e-mail: hejialing@cqu.edu.cn, txiang@cqu.edu.cn).}
        
    \thanks{Geng Sun is with College of Computer Science and Technology, Jilin University, China 130012, (e-mail: sungeng@jlu.edu.cn).}

    \thanks{Z. Xiong is with the School of Electronics, Electrical Engineering and Computer Science (EEECS), Queen's University Belfast, Belfast, BT7 1NN, U.K. (z.xiong@qub.ac.uk).}

    
    \thanks{S. Mao is with the Department of Electrical and Computer Engineering, Auburn University, Auburn, USA (e-mail: smao@ieee.org).}
    
    \thanks{Dong In Kim is with the Department of Electrical and Computer Engineering, Sungkyunkwan University, Suwon 16419, South Korea (email: dongin@skku.edu).}
    }

\maketitle

\begin{abstract}

The increasing saturation of terrestrial resources has driven the exploration of low-altitude applications such as air taxis. Low altitude wireless networks (LAWNs) serve as the foundation for these applications, and integrated sensing and communication (ISAC) constitutes one of the core technologies within LAWNs. However, the openness nature of low-altitude airspace makes LAWNs vulnerable to malicious channel access attacks, which degrade the ISAC performance. Therefore, this paper develops a game-based framework to mitigate the influence of the attacks on LAWNs. Concretely, we first derive expressions of communication data's signal-to-interference-plus-noise ratio and the age of information of sensing data under attack conditions, which serve as quality of service metrics. Then, we formulate the ISAC performance optimization problem as a Stackelberg game, where the attacker acts as the leader, and the legitimate drone and the ground ISAC base station act as second and first followers, respectively. On this basis, we design a backward induction algorithm that achieves the Stackelberg equilibrium while maximizing the utilities of all participants, thereby mitigating the attack-induced degradation of ISAC performance in LAWNs. We further prove the existence and uniqueness of the equilibrium. Simulation results show that the proposed algorithm outperforms existing baselines and a static Nash equilibrium benchmark, ensuring that LAWNs can provide reliable service for low-altitude applications.

\end{abstract}

\vspace{-0.35cm}

\IEEEpeerreviewmaketitle

\section{Introduction}

As ground resources become increasingly constrained, the focus of economic development is gradually shifting from the ground to the low-altitude airspace~\cite{yuan2025ground}. In this domain, various applications, such as low-altitude logistics and urban air taxi, are emerging~\cite{sun2025recovery}. For instance, drones can expedite parcel deliveries, while autonomous electric vertical take-off and landing (eVTOL) vehicles can service as the air taxi to relieve stress of ground transportation~\cite{xiong2024evtol}. To support these low-altitude applications, low-altitude wireless networks (LAWNs) become a crucial enabling infrastructure. LAWNs provide the communication, computing, and control capability needed to coordinate the operations of the low-altitude drones~\cite{jin2025predictive}. In essence, LAWNs connect low-altitude vehicles to ground base stations (BSs) and with each other, thereby enabling real-time data exchange, decision making, and remote control. 

One of the core technologies in LAWNs is integrated sensing and communication (ISAC), which can simultaneously provide data exchange and environmental sensing functions~\cite{wang2024generative}. This indicates the transmitted signals can not only carry information, but also act as probes that detect obstacles or track time-sensitive targets. For example, by sending out the wireless signal, a ground ISAC BS can communicate with drones and scan the surroundings to collect sensing data~\cite{mu2023uav}, which can enable quick reactions for applications such as air traffic management. Moreover, drones equipped with reconfigurable intelligent surfaces (RIS) can even dynamically reflect and shape signals, thereby extending coverage and supporting more low-altitude applications~\cite{jiang2025integrated}.

While appealing, the openness of the low-altitude airspace exposes LAWNs to significant security threats. A particular threat is the channel access attack~\cite{wei2020classification}, wherein an attacker can jam the wireless channels, thereby preventing drones from receiving updates, increasing the risk of air traffic accidents. In response, researchers have developed various defense strategies. In~\cite{wei2023integrated}, a drone recycles artificial noise for joint jamming and sensing, while Kalman filtering with a neural network predictor adapts its trajectory and resources to deter eavesdropping. Beyond waveform-level defenses, researchers have developed secure protocols to protect data integrity and confidentiality in adversarial settings~\cite{khan2022secure}, designed robust channel-access strategies that sustain reliable detection and data exchange under jamming~\cite{yang2023age}, and applied blockchain-based encryption schemes to strengthen the security and privacy of drone data \cite{ch2020security}. These studies assume a static and nonadaptive attacker. In practice, however, an attacker can dynamically vary its strategy by switching frequencies or adjusting jamming power. Ignoring this adaptability makes it difficult to analyze the hierarchical and iterative game between the attacker and LAWNs and leads to weaker defenses.

To address these challenges, this paper proposes a defense framework by modeling the interactions between the channel access attacker and LAWNs as a Stackelberg game to optimize the ISAC performance. In this game, the attacker acts as the leader by first selecting its attack strategy, while the  LAWNs' operators, including the legitimate drone with RIS and the ground ISAC base station, serve as the follower that optimally adjusts its response. Such a framework reflects the reality of adaptive threats, i.e., the attacker launches the attack first and subsequently LAWNs optimize the defense strategy given the attacker’s actions. By analyzing this interaction, we derive the Stackelberg equilibrium, which jointly maximizes the attacker’s utility and the LAWNs' sensing and communication performance. Compared with existing static methods, the proposed framework allows the LAWN to proactively fine-tune its transmission and sensing strategies in anticipation of an adaptive adversary, thereby maintaining ISAC performance under dynamic channel-access attacks. The main contributions of this paper are summarized below.


\begin{itemize}
\item We propose an age of information (AoI)-based metric to describe the freshness of sensing data in LAWNs. This metric inherently captures the temporal dynamics and the freshness of the information about the state of a time-sensitive target, crucial for low-altitude applications.

\item We formulate a Stackelberg game-based ISAC performance optimization problem for time-sensitive target under channel access attacks, where the attacker, legitimate drone, and ISAC BS act as the game leader, second follower, and first follower, respectively. Such modeling analyzes the hierarchical confrontation relationship and game iteration process between the attacker and LAWNs, making LAWNs capable of handling dynamic attacks.

\item We design a backward induction (BI)-based algorithm to solve the optimization problem. We further theoretically prove the existence and uniqueness of the Stackelberg equilibrium for the formulated game and show that the equilibrium maximizes the utilities of all players.

\item Simulations are conducted under different parameter settings and the results show that the proposed algorithm outperforms the baselines, while reaching higher utilities than the Nash equilibrium.
\end{itemize}

The paper is organized as follows. Section~\ref{RW} reviews related work. Section~\ref{SM} gives the system model. Section~\ref{IP} derives the communication signal to interference plus noise ratio (SINR) and the freshness of sensing data. Section~\ref{OP} formulates the Stackelberg game and presents the solution. Section~\ref{EA} reports evaluation results. Section~\ref{CC} concludes the paper.

\section{Related work} \label{RW}

\subsection{ISAC Performance Optimization in LAWNs}
ISAC enables both wireless sensing and communication using the shared network resources, which can improve LAWNs efficiency~\cite{wang2025generative}. In \cite{chen2025full}, the authors modeled the drone as a low altitude platform and adopted an alternating optimization scheme that jointly optimizes task allocation, computing resources, and transmit and receive beamforming, reducing system energy consumption by up to 54.12\%. The authors in~\cite{meng2022throughput} maximized the drone's achievable rate with the sensing frequency and beam pattern gain constraints. By jointly optimizing drone trajectory, user association, target sensing selection, and transmit beamforming, the system's sum rate can reach to 12 bps/Hz. Besides, the authors in~\cite{jing2024isac} formulated the drone trajectory design as a weighted optimization problem and propose a multi-stage optimization algorithm to obtain the optimal while optimizing ISAC performance.


From the antenna design perspective, the authors in~\cite{kuang2024movable} proposed a movable-antenna array-enabled ISAC system for low-altitude applications, which adjusts the antenna array's positions via the alternation optimization to improve achievable data rate and beamforming gain. In~\cite{kuang2024movable}, the authors design an ISAC system for low-altitude scenarios that employs a movable antenna array. By alternately optimizing the array’s positions, the system raises both the achievable data rate and the beamforming gain. The authors in~\cite{ye2024integrated} focus on signal design and develop a low-altitude ISAC system where deep reinforcement learning (DRL) jointly optimizes BS beamforming and drone trajectories, enhancing the communication sum rate while satisfying sensing constraints. As can be seen, current studies optimize ISAC in LAWNs through resource allocation, UAV trajectory planning, and related methods~\cite{tang2025cooperative}. However, the openness of the low altitude airspace exposes sensing and communication to wireless attacks, so improving ISAC performance under attack is essential.


\subsection{Game-Theoretic Defense of ISAC in LAWNs}
Game theory models the strategic interaction between attackers and defenders, providing a foundation for designing resilient ISAC in LAWNs. For instance, in~\cite{liu2024game}, the authors formulate a Bayesian Stackelberg game where the BS (leader) optimizes the transmit precoder and the jammer (follower) selects its power. Using semidefinite relaxation with Gaussian randomization, they derive closed‑form strategies, and simulations show the Stackelberg equilibrium yields higher utility than baseline methods. To safeguard the ISAC BS from a mobile hostile drone, the authors in~\cite{mamaghani2025securing} constructed a non-cooperative Stackelberg game and proposed a DRL algorithm to jointly refine resource allocation and drone trajectory, thereby enhancing secrecy rate, radar accuracy, and energy efficiency. In~\cite{chen2024anti}, the authors tackle resource allocation under composite jamming and interchannel interference by embedding game theory into a DRL framework, which maximizes the communication rate and sensing power, outperforming baseline methods on both sensing and communication metrics.


For drone-aided traffic monitoring, the authors in~\cite{yang2022aoi} formulated a three-layer Stackelberg game that jointly determines sensing rate, transmission power, and attack power, attaining an age-of-information (AoI) minimum via sub-gradient updates. In multi-channel IoT networks subject to dynamic access attacks, the authors in~\cite{yang2021game} recast channel selection as two ordinary potential games, prove the existence of Nash equilibrium, and design a reinforcement-learning algorithm that converges to them, yielding lower AoI. To support personalized anti-jamming in heterogeneous drone swarms, the authors in~\cite{qin2025multi} modeled the channel-and-power selection problem as an adversarial game with a unique equilibrium and propose a personalized federated soft actor-critic algorithm that markedly improves interference resilience. As can be seen, prior studies have shown that game theoretic defenses are effective across wireless networks~\cite{feng2019dynamic}. Hence, we cast ISAC performance optimization for time-sensitive targets under channel access attack as a Stackelberg game, enabling LAWNs to provide secure and efficient services.

\section{System model}\label{SM}

\subsection{Low altitude economy network model}

\begin{figure}[t]
  \centering
  \includegraphics[width=8cm]{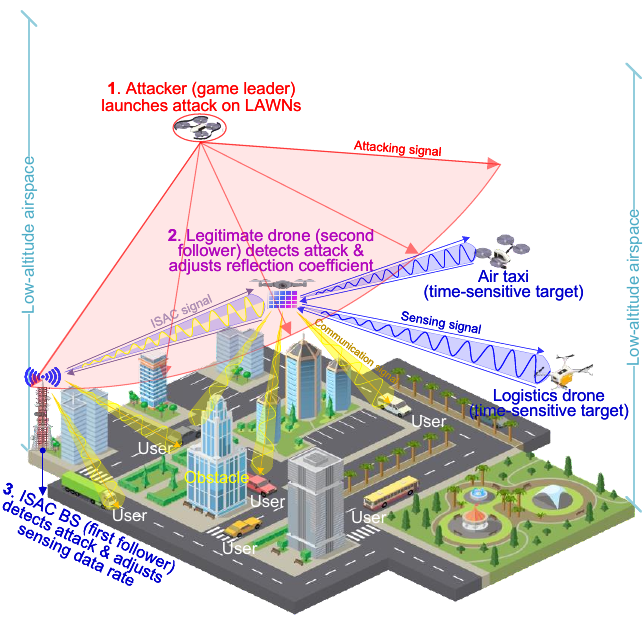}\\
  \caption{System model of the LAWNs under channel access attack. Here, the malicious UAV first launches an attack by injecting additional noise to the wireless channel. In response, the legitimate UAV with RIS and BS respectively adjust the gain and sensing data generation rate to mitigate the impact of the attack, thereby maintaining the ISAC performance.}
  \vspace{-0.3cm}
  \label{SYSTEMMODEL}
\end{figure}

Figure \ref{SYSTEMMODEL} presents the low-altitude scenario supported by LAWNs, which includes the dual-function ISAC BS, mobile users, legitimate drone equipped with RIS for communication and sensing, time-sensitive targets, and the illegitimate drone for channel access attack. Specifically, the ISAC BS is equipped with multiple antennas that support communication with multiple users and drones while transmitting sensing waveforms to detect nearby targets. Legitimate low-altitude drones equipped with RIS can communicate with the BS and relay both communication and sensing services to targets beyond the base station’s reach. In parallel, we consider a malicious drone that executes a channel-access attack aimed at degrading the ISAC performance of the LAWNs.
\vspace{-0.2cm}
\subsection{ISAC Signal Model at BS}
Let $\left( {{x_{BS}},{y_{BS}},{z_{BS}}} \right)$ be the location of the BS, $M$ represents the number of transmitting antenna of BS, and ${\mathbf{s} {_r}} \in {{\mathbb{C}}^{M \times 1}}$ and ${\mathbf{s}{_c}} \in {{\mathbb{C}}^{N \times 1}}$ be the sensing signal and communication symbol to $N$ users, respectively. Therefore, we denote ${{\bf{B}}_{\rm r}} = [{{\mathop{\rm b}\nolimits} _{{\mathop{\rm r}\nolimits} ,1}},{{\mathop{\rm b}\nolimits} _{{\mathop{\rm r}\nolimits} ,2}}, \ldots ,{{\mathop{\rm b}\nolimits} _{{\mathop{\rm r}\nolimits} ,M}}] \in {{\mathbb{C}}^{M \times M}}$ and ${{\bf{B}}_{\rm c}} = [{{\mathop{\rm b}\nolimits} _{{\mathop{\rm c}\nolimits} ,1}},{{\mathop{\rm b}\nolimits} _{{\mathop{\rm c}\nolimits} ,2}}, \ldots ,{{\mathop{\rm b}\nolimits} _{{\mathop{\rm c}\nolimits} ,N}}] \in {{\mathbb{C}}^{M \times N}}$ as the beamforming matrices for sensing and communication signals, and then the transmitted signal of BS can be expressed as:
\begin{equation}\label{eq1}
{\mathop{\bf{s}}\nolimits}  = {{\bf{B}}_{\rm r}}{\mathbf{s}_{\rm r}} + {{\bf{B}}_{\rm c}}{\mathbf{s}_{\rm c}} = [{{\bf{B}}_{\rm r}},{{\bf{B}}_{\mathop{\rm c}\nolimits} }]{[{\left( {{{\mathop{\mathbf{s}}\nolimits} _{\rm r}}} \right)^{\mathop{\rm T}\nolimits} },{\left( {{{\mathop{\mathbf{s}}\nolimits} _{\rm c}}} \right)^{\mathop{\rm T}\nolimits} }]^{\mathop{\rm T}\nolimits} } = {\bf{B}}\hat {\mathbf{s}},
\end{equation}
where ${\bf{B}} \in {{\mathbb{C}}^{M \times \left( {M + N} \right)}}$ is the ISAC transmission beamforming matrix. Without loss of generality, according to~\cite{liu2020joint}, the sensing signal can be generated through the pseudo-random coding with ${\mathbb{E}}\left[ {\mathop{\mathbf{s}}\nolimits}  \right] = {\bf{0}}$ and ${\mathbb{E}}\left[ {{\mathop{\mathbf{s}}\nolimits} {{\mathop{\mathbf{s}}\nolimits} ^{\rm T}}} \right] = {{\bf{I}}_M}$. On this basis, let the transmission symbol ${{\mathop{\mathbf{s}}\nolimits} _{\rm c}}$ satisfies $\mathcal{CN}\left( {{\bf{0}},{{\bf{I}}_K}} \right)$, then the covariance matrix of the transmitted signal is determined as:
\begin{equation}\label{eq2}
\mathbf{R} = \mathbb{E}\left[ \rm {{\mathbf{s}{\mathbf{s}^H}}} \right] = \mathbf{B}{\mathbf{B}^{\rm H}} = {\mathbf{B}_{\rm r}}\mathbf{B}_{\rm r}^{\rm H} + \sum_{i=1}^{N} {\mathbf{R}}_i,
\end{equation}
where ${\mathbf{R}}_i$ is the rank-1 matrix, calculated as ${{\mathbf{R}}_i} = {{\rm b}_{{\rm c},i}}{\rm b}_{{\rm c},i}^{\rm H}$.
\subsection{Wireless Channel Model}
Let the RIS of legitimate drone has $P$ reflecting elements, then the wireless channel between BS and drone, drone and the $i$-th user, and BS and the $i$-th user, can be denoted as ${\bf{H}} \in {{\mathbb{C}}^{P \times M}}$, ${{\bf{h}}_{1,i}} \in {{\mathbb{C}}^{P \times 1}}$, and ${{\bf{h}}_{2,i}} \in {{\mathbb{C}}^{M \times 1}}$, respectively. On this basis, the BS receives the wireless signals and estimates the channel state information (CSI). In the considered LAWNs, the communication link can be modeled as the large-scale fading channel, and the path loss can be calculated as:
\begin{equation}\label{eq4}
Loss\left[ {dB} \right] = 10{\beta _1}{\log _{10}}\left( {\frac{{{r_s}}}{{{r_0}}}} \right) + {\beta _2} + 10{\beta _3}{\log _{10}}\left( {\frac{{{f_s}}}{{{f_0}}}} \right) + {\beta _4},
\end{equation}
where $\beta _1$ and $\beta _3$ are the coefficients indicating the impact of path loss on distance and frequency aspects, respectively, $\beta _2$ is the optimized offset for path loss, $r_s$ and $f_s$ represent the distance from the transmitter to receiver and the carrier frequency, respectively, and $r_0 = 1$ m and $f_0 = 1 $ GHz are the reference for distance and frequency metrics, respectively. Moreover, $\beta_4$ denotes the signal's large-scale fluctuations due to random shadowing on the mean path loss over distance.

Given the communication users are distributed with densely surrounded scatters, the channel (${{\bf{h}}_{1,i}}$) between drone and the $i$-th user, and the channel (${{\bf{h}}_{2,i}}$) between BS and the $i$-th user can be represented by the Rayleigh fading channels. Meanwhile, as the link between the BS and the drone comprises both line-of-sight (LoS) and non-line-of-sight (NLoS) components, then the corresponding channel (${\bf{H}}$) between BS and drone can be represented by the Rician fading channel. More concretely, for the Rician fading channel, the NLoS component ${\mathbf{H}_{NLoS}}$ follows the complex normal distribution with zero-mean vector and the covariance that is the Kronecker product of $\Sigma _ {BS} $ and $\Sigma _ {RIS} $, indicating 
\begin{equation}\label{eq3}
{{\bf{H}}_{NLoS}} \sim \mathcal{CN}\left( {\mathbf{0},\Sigma _ {RIS}  \otimes \Sigma _ {BS} } \right).
\end{equation}
Here, $\Sigma _{BS}\geq 0$ and $\Sigma _{RIS}\geq 0$ represent the spatial correlation matrices with unit diagonal elements at BS and drone, respectively. Additionally, the LoS part ${\mathbf{H}_{LoS}}$ in the Rician fading channel can be defined as 
\begin{equation}\label{eq5}
{{\bf{H}}_{LoS}} = {{\mathop{\rm a}\nolimits} _2}\left( {{\theta _2}} \right){\mathop{\rm a}\nolimits} _1^{\mathop{\rm H}\nolimits} \left( {{\theta _1}} \right),
\end{equation}
where $\theta_1$ is the the angle of departure (AoD) at the BS, $\theta_2$ is the angle of arrival (AoA) at the drone. Besides, ${{\mathop{\rm a}\nolimits} _1}\left( {{\theta _1}} \right)$ and ${{\mathop{\rm a}\nolimits} _2}\left( {{\theta _2}} \right)$ represent the array response vectors of the antennas from BS and drone, which are expressed as
\begin{equation}\label{eq6}
{{\mathop{\rm a}\nolimits} _1}\left( {{\theta _1}} \right) = {\left[ {1,{e^{ - j2\pi {d_{BS}}\frac{{\sin \left( {{\theta _1}} \right)}}{\omega }}}, \ldots ,{e^{ - j2\pi {d_{BS}}\left( {M - 1} \right)\frac{{\sin \left( {{\theta _1}} \right)}}{\omega }}}} \right]^{\mathop{\rm T}\nolimits} }
\end{equation}
and
\begin{equation}\label{eq7}
{{\mathop{\rm a}\nolimits} _2}\left( {{\theta _2}} \right) = {\left[ {1,{e^{ - j2\pi {d_{RIS}}\frac{{\sin \left( {{\theta _2}} \right)}}{\omega }}}, \ldots ,{e^{ - j2\pi {d_{RIS}}\left( {P - 1} \right)\frac{{\sin \left( {{\theta _2}} \right)}}{\omega }}}} \right]^{\mathop{\rm T}\nolimits} },
\end{equation}
respectively, where $d_{BS}$ represents the antenna spacing at the BS, $d_{RIS}$ is the element spacing distance of the RIS at drone, and $\omega $ is the wavelength of the transmitted signal. For ease of computation, as well as the AoA and AoD estimations, both $d_{BS}$ and $d_{RIS}$ are set to ${\omega  \mathord{\left/
 {\vphantom {\omega  2}} \right.
 \kern-\nulldelimiterspace} 2}$.

As for the target sensing aspect, similar to (\ref{eq6}) and (\ref{eq7}), let $\theta_3$ be the AoA of the RIS, then the channel response matrix between drone and the target can be obtained
\begin{equation}\label{eq8}
{\bf{T}} = {\beta _5}{{\mathop{\rm a}\nolimits} _3}\left( {{\theta _3}} \right){\mathop{\rm a}\nolimits} _3^{\mathop{\rm H}\nolimits} \left( {{\theta _3}} \right) \in {{\mathbb{C}}^{P \times P}},
\end{equation}
where the amplitude $\beta_5$ is the multiplicative fading of the drone-based target sensing channel. Based on \cite{yu2023active}, the target acts as a single scattering object and the its loss and gain factors are negligible. Hence, the energy of the sensing signal reflected from the RIS and received at the BS is calculated as:
\begin{equation}\label{eq9}
{P_{sense}} = {{{P_{trans}}{G^2}{\omega ^2}S} \mathord{\left/
 {\vphantom {{{P_{trans}}{G^2}{\omega ^2}S} {{{\left( {4\pi } \right)}^3}{R^4}}}} \right.
 \kern-\nulldelimiterspace} {{{\left( {4\pi } \right)}^3}{R^4}}},
\end{equation}
where $P_{trans}$ and $G$ are the transmission power and antenna gain for sending the sensing signal, respectively, $S$ is the target's radar cross section (RCS), and $R$ represents the distance between drone and target. By treating the drone as the monostatic multiple-input and multiple-output radar, the path loss of the drone-based sensing link ($\beta_5$) can be calculated as:
\begin{equation}\label{eq10}
{\beta _5} = \sqrt {{{{\omega ^2}S} \mathord{\left/
 {\vphantom {{{\omega ^2}S} {{{\left( {4\pi } \right)}^3}{R^4}}}} \right.
 \kern-\nulldelimiterspace} {{{\left( {4\pi } \right)}^3}{R^4}}}}  .
\end{equation}

\subsection{Legitimate Drone Model}
For the RIS deployed on the drone, its reflecting coefficient matrix can be denoted as ${\bf{\Phi }} = {\mathop{\rm Diag}\nolimits} \left( {\left[ {{g_1}, \ldots ,{g_i}, \ldots ,{g_P}} \right]} \right)$. Note that due to the limited energy supply of the drone, we let the power amplification gain of all the RIS's reflecting elements to be the same, indicating ${\left| {{g_1}} \right|^2} =  \cdots  = {\left| {{g_P}} \right|^2} = { {{g}}^2}$. In addition, the maximum power amplification gain of the reflecting element is denoted by $g_{max}$, and the RIS position is represented by the coordinates $\left( {{x_{RIS}},{y_{RIS}},{z_{RIS}}} \right)$. On this basis, as presented in Fig. \ref{SYSTEMMODEL}, the RIS facilitates two kinds of signal reflections. First, it redirects the signals transmitted by the BS toward both the users (for communication) and the time-sensitive target (for sensing). Then, the echo signal returning from the time-sensitive target is reflected back to the BS. According to \cite{liu2021active}, the above reflections is:
\begin{equation}\label{eq11}
{\bf{r}}_1^{refle} = {\bf{\Phi Hs}} + {\bf{\Phi }}{{\bf{n}}_1},
\end{equation}
\begin{equation}\label{eq12}
{\bf{r}}_2^{refle} = {{\bf{\Phi }}^{\mathop{\rm H}\nolimits} }{\bf{T\Phi Hs}} + {{\bf{\Phi }}^{\mathop{\rm H}\nolimits} }{\bf{T\Phi }}{{\bf{n}}_1} + {{\bf{\Phi }}^{\mathop{\rm H}\nolimits} }{{\bf{n}}_2},
\end{equation}
where ${{\bf{n}}_1} \sim \mathcal{CN}\left( {{\bf{0}},{\sigma ^2}{{\bf{I}}_K}} \right)$ and ${{\bf{n}}_2} \sim \mathcal{CN}\left( {{\bf{0}},{\sigma ^2}{{\bf{I}}_K}} \right)$ are the additive white Gaussian noise (AWGN) at the RIS end with the same distribution of the noise power $\sigma^2$. 

\subsection{Channel Access Attack Drone Model}
To attack various applications supported by LAWNs, the malicious drone first launches the channel access attack \cite{wei2020classification}, that is spoofing the media access control (MAC) protocol and then invading the communication and sensing channel. After obtaining access, it further adds the extra noise into the channels, including the channels between BS and drone, drone and user, BS and user, and drone and target, thereby degrading the ISAC performance of LAWNs. Moreover, to reduce the probability of being detected, we let the added noise of the attacker, denoted as ${{\bf{n}}_{att}}$, have the same distribution as the AWGN, whose power density is $\sigma_{att}^2$ with the maximum value of $\sigma^2$. For example, for the channel from drone to the users, after the channel access attack is launched, the attacker adds the noise with distribution of ${{\bf{n}}_{att}} \sim \mathcal{CN}\left( {{\bf{0}},{\sigma_{att} ^2}{{\bf{I}}_K}} \right)$.

\section{ISAC Performance of Time-sensitive Target under Attack}\label{IP}

\subsection{Communication SINR Analysis }
With the help of an RIS on the drone, the users can receive communication signals from two propagation paths, including the direct link from the BS to users, as well as the reflected link from drone to users. Building on this, the signal received by the $i$-th user can be expressed as:
\begin{equation}\label{eq13}
{\bf{r}}_{c,i}^{com} = {\bf{h}}_{1,i}^{\mathop{\rm H}\nolimits} {\bf{\Phi Hs}} + {\bf{h}}_{1,i}^{\mathop{\rm H}\nolimits} {\bf{\Phi }}{{\bf{n}}_1} + {\bf{h}}_{2,i}^{\mathop{\rm H}\nolimits} {\bf{s}} + {{\mathop{\rm y}\nolimits} _i},
\end{equation}
where ${{\mathop{\rm y}\nolimits} _i}$ is the AWGN at the user with variance of $\sigma _y^2$. Considering the channel access attack, after defining the equivalent channel between BS and the $i$-th user as ${\bf{h}}_i^{\mathop{\rm H}\nolimits}  = {\bf{h}}_{1,i}^{\mathop{\rm H}\nolimits} {\bf{\Phi H}} + {\bf{h}}_{2,i}^{\mathop{\rm H}\nolimits} $, then the SINR of the $i$-th user can be evaluated as
\begin{align}\label{eq14}
&\mathrm{SINR}_i^{\mathrm{com}} = \\ \notag
&\frac{\mathbf{h}_i^{\mathrm{H}} \mathbf{R}_i \mathbf{h}_i}{\mathbf{h}_i^{\mathrm{H}} \left( \mathbf{R} - \mathbf{R}_i \right) \mathbf{h}_i + \left( \sigma^2 + \sigma_{\mathrm{att}}^2 \right) \mathbf{h}_{1,i}^{\mathrm{H}} \mathbf{\Phi} \mathbf{\Phi}^{\mathrm{H}} \mathbf{h}_{1,i} + \left( \sigma_{\mathrm{att}}^2 + \sigma_y^2 \right)}.
\end{align}

\subsection{Freshness of Sensing Data}
In the considered low-altitude airspace, for tracking the time-sensitive target, the SINR metric primarily reflects the instantaneous quality of the received signal for a single measurement. A higher SINR indicates a stronger and better signal at a specific moment. However, it does not inherently capture the temporal dynamics and the freshness of the information about the target's state. Such a state is crucial, especially for those applications that require fast response, such as an air taxi. To quantify information freshness, the AoI metric is required. Hence, in this section, we analyze sensing data freshness by deriving closed-form AoI expressions using queuing theory.

\subsubsection{Transmission of Sensing Data}
During sensing, the BS receives the echo signals reflected from RIS of drone, which contain three main components, including the thermal noise produced at both the BS and drone, the self-interference term that propagates along the BS–drone–BS path, and the desired sensing signals that travels via the BS–drone–target–drone–BS route. As the BS–drone–BS component conveys no information about the target, it is treated like interference. Hence, the aggregated signals at the BS can be expressed as:
\begin{equation}\label{4511111}
\begin{split}
\mathbf{\tilde r}^{\mathrm{BS}} &= \mathbf{H}^{\mathrm{H}} \left( \mathbf{r}_1^{\mathrm{refle}} + \mathbf{r}_2^{\mathrm{refle}} \right) + {{\mathbf{w}}_r} \\
&= \mathbf{H}^{\mathrm{H}} \mathbf{\Phi}^{\mathrm{H}} \mathbf{T} \mathbf{\Phi} \mathbf{H} \mathbf{s} + \mathbf{H}^{\mathrm{H}} \mathbf{\Phi}^{\mathrm{H}} \mathbf{T} \mathbf{\Phi} \mathbf{n}_1 + \mathbf{H}^{\mathrm{H}} \mathbf{\Phi} \mathbf{n}_1 \\
&+ \mathbf{H}^{\mathrm{H}} \mathbf{\Phi}^{\mathrm{H}} \mathbf{n}_2 + \mathbf{H}^{\mathrm{H}} \mathbf{\Phi} \mathbf{H} \mathbf{s} + {{\mathbf{w}}_r},
\end{split}
\end{equation}
where ${{\bf{H}}^{\mathop{\rm H}\nolimits} }{\bf{\Phi Hs}}$ is the interference echo, and ${{\mathbf{w}}_r}$ is the AWGN at the BS end, ${\sigma _r^2}$ is the noise power density, and ${{\mathbf{w}}_r}\sim \mathcal{CN}\left( {{\bf{0}},\sigma _r^2{{\bf{I}}_M}} \right)$. With the aid of the self-interference (SI) cancellation, the interference echo can be alleviated. Therefore, in the presence of the channel access attack, the received sensing signal can be further expressed as

\begin{equation}\label{9265561232}
\begin{split}
\mathbf{r}^{\mathrm{BS}} &= \mathbf{H}^{\mathrm{H}} \mathbf{\Phi}^{\mathrm{H}} \mathbf{T} \mathbf{\Phi} \mathbf{H} \mathbf{s} + \mathbf{H}^{\mathrm{H}} \mathbf{\Phi}^{\mathrm{H}} \mathbf{T} \mathbf{\Phi} \mathbf{n}'_1 + \mathbf{H}^{\mathrm{H}} \mathbf{\Phi} \mathbf{n}'_1 \\
&\quad + \mathbf{H}^{\mathrm{H}} \mathbf{\Phi}^{\mathrm{H}} \mathbf{n}'_2 + \varepsilon \mathbf{H}^{\mathrm{H}} \mathbf{\Phi} \mathbf{H} \mathbf{s} + {{\mathbf{w'}}_r},
\end{split}
\end{equation}
where $\varepsilon$ represents the SI coefficient, ${{\bf{n}'}_1} \sim \mathcal{CN}\left( {{\bf{0}},{{\left( {{\sigma^2} + {\sigma^2 _{att}}} \right)}}{{\bf{I}}_K}} \right)$, ${{\bf{n}'}_2} \sim \mathcal{CN}\left( {{\bf{0}},{{\left( {{\sigma^2} + {\sigma^2 _{att}}} \right)}}{{\bf{I}}_K}} \right)$, and ${{{\bf{w'}}}_r} \sim \mathcal{CN}\left( {{\bf{0}},{{\left( {{\sigma^2 _r} + {\sigma^2 _{att}}} \right)}}{{\bf{I}}_M}} \right)$. On this basis, the interference-plus-noise covariance matrix is calculated as:
\begin{equation}\label{11531321}
{\bf{J}} = {\bf{Z}} + {\bf{CR}}{{\bf{C}}^{\mathop{\rm H}\nolimits} },
\end{equation}
where 
\begin{align}\label{eq18}
{\bf{Z}} &= {\left( {\sigma^2  + {\sigma _{att}^2}} \right)}\left( {{{\bf{H}}^{\mathop{\rm H}\nolimits} }{\bf{\Phi }}{{\bf{\Phi }}^{\mathop{\rm H}\nolimits} }{\bf{T\Phi H}} + {{\bf{H}}^{\mathop{\rm H}\nolimits} }{{\bf{\Phi }}^{\mathop{\rm H}\nolimits} }{\bf{T\Phi }}{{\bf{\Phi }}^{\mathop{\rm H}\nolimits} }{\bf{H}}} \right) \\ \notag
&+ {\left( {\sigma^2  + {\sigma _{att}^2}} \right)}{\bf{X}}{{\bf{X}}^{\mathop{\rm H}\nolimits} }  + 2{\left( {\sigma^2  + {\sigma _{att}^2}} \right)}{{\bf{H}}^{\mathop{\rm H}\nolimits} }{{\bf{\Phi }}^{\mathop{\rm H}\nolimits} }{\bf{\Phi H}} \\ \notag
&+ {\left( {{\sigma _r^2} + {\sigma _{att}^2}} \right)}{{\bf{I}}_M},
\end{align}
representing the equivalent noise covariance matrix, ${\bf{X}} = {{\bf{H}}^{\mathop{\rm H}\nolimits} }{{\bf{\Phi }}^{\mathop{\rm H}\nolimits} }{\bf{T\Phi }} \in {{\mathbb{C}}^{M \times P}}$, and ${\bf{C}} = \varepsilon {{\bf{H}}^{\mathop{\rm H}\nolimits} }{\bf{\Phi H}}$. On this basis, by defining ${\bf{F}} = {{\bf{H}}^{\mathop{\rm H}\nolimits} }{{\bf{\Phi }}^{\mathop{\rm H}\nolimits} }{\bf{T\Phi H}} \in {{\mathbb{C}}^{M \times M}}$, the SINR of the sensing signal can be obtained
\begin{equation}\label{48453123132}
{{\mathop{\rm SINR}\nolimits} ^{sense}} = {\mathop{\rm Tr}\nolimits} \left( {{\bf{FR}}{{\bf{F}}^{\mathop{\rm H}\nolimits} }{{\bf{J}}^{ - 1}}} \right),
\end{equation}
where ${\rm{Tr}}\left(  \cdot  \right)$ is the trace calculator. Finally we let $\eta $ represent the bandwidth of the BS, then the transmission rate of the sensing signal can be evaluated as
\begin{equation}\label{651231}
{\gamma _{sense}} = \eta {\log _2}\left( {1 + {{{\mathop{\rm SINR}\nolimits} }^{sense}}} \right).
\end{equation}

\subsubsection{Average AoI of Sensing Data}
\begin{figure}[t]
  \centering
  \includegraphics[width=8.5cm]{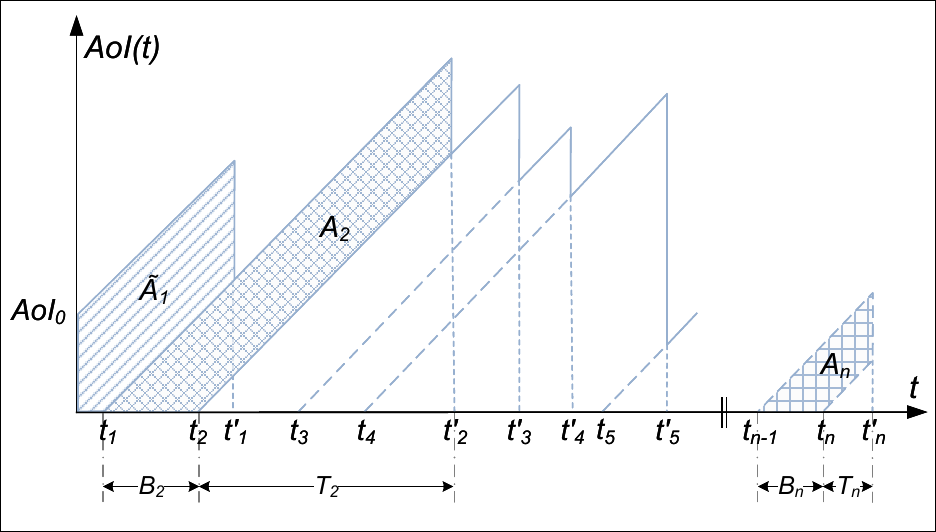}\\
  \caption{The modeling of the AoI metric.}
  \vspace{-0.3cm}
  \label{AoI}
\end{figure}

\
As shown in Fig.~\ref{AoI}, the AoI curve can be represented by the sawtooth function, where the sensing data AoI first increases with time increasing. When the sensing data is successfully received by the BS, the AoI value decreases immediately. Based on queuing theory, the sensing data generated by the BS can be regarded as the customer, while the wireless links (from BS to drone, drone to target, target to drone, and drone to BS) can act as the server. Let $AoI_0$ be the initial AoI value of sensing data at $t=0$, $t_i$ be the generation time of the $i$-th sensing data, and $t'_i$ be the time that $i$-th sensing data is received by terminal end. Building on this, at $t'_i$, the AoI of the sensing data can be expressed as $T_i = {{t'}_i} - {t_i}$. Note that $T_i$ represents the system time of the $i$-th sensing data, including the time waiting for service and the time being served. Therefore, within the time interval $\left( {0,\tau } \right)$, the average AoI of the sensing  data can be calculated as:
\begin{equation}\label{7856231335}
AAo{I_\tau } = \frac{1}{\tau }\int_0^\tau  {AoI\left( t \right)} \,dt.
\end{equation}
For ease of presentation, the average AoI can be further calculated as the unit area of the graph enclosed by the sawtooth function and time axes, which can be denoted as:
\begin{equation}\label{2323135}
AAo{I_\tau } = \frac{1}{\tau }\left( {{{\tilde A}_1} + \sum\nolimits_{i = 2}^{\Omega \left( \tau  \right)} {{A_i}}  + \frac{{T_n^2}}{2}} \right),
\end{equation}
where ${{{\tilde A}_1}}$ is the polygon area, ${A_i}\left( {2 \le i \le n} \right)$ is the trapezoid areas, ${{T_n^2} \mathord{\left/
 {\vphantom {{T_n^2} 2}} \right.
 \kern-\nulldelimiterspace} 2}$ is the triangular area, $\Omega \left( \tau  \right) = \max \left\{ {n|{t_n} \le \tau } \right\}$ defines the number of arrivals by time $\tau$. Here, ${A_i}\left( {2 \le i \le n} \right)$ can be calculated as the difference of two isosceles triangles. Let $B_i = t_i-t_{i-1}$ denote the elapsed time between the producing of $i$-th and $(i-1)$-th sensing data, then we have:
\begin{equation}\label{12545323}
{A_i} = \frac{1}{2}{\left( {{T_i} + {B_i}} \right)^2} - \frac{{T_i^2}}{2} = {B_i}{T_i} + \frac{{B_i^2}}{2}.
\end{equation}

By regarding the generation of the sensing data as a stochastic process, $B_i$ can represent the inter-arrival time of the generation of the $i$-th sensing. Therefore, (\ref{2323135}) can be further expressed as:
\begin{equation}\label{12465523}
AAo{I_\tau } = \frac{{\tilde A}}{\tau } + \frac{{\Omega \left( \tau  \right) - 1}}{\tau }\frac{1 }{{\Omega \left( \tau  \right) - 1}}\sum\nolimits_{i = 2}^{\Omega \left( \tau  \right)} {\left[ {{B_i}{T_i} + \frac{{B_i^2}}{2}} \right]},
\end{equation}
where $\tilde A = {\tilde A_1} + {{T_n^2} \mathord{\left/
 {\vphantom {{T_n^2} 2}} \right.
 \kern-\nulldelimiterspace} 2}$. Note that when $\tau  \to \infty $, ${{\tilde A} \mathord{\left/
 {\vphantom {{\tilde A} \tau }} \right.
 \kern-\nulldelimiterspace} \tau } \to 0$. In this way, let $\lambda$ denote the generation rate of sensing data, we have
\begin{equation}\label{9656664}
\lambda  = \mathop {\lim }\limits_{\tau  \to \infty } \left[ {{{\Omega \left( \tau  \right)} \mathord{\left/
 {\vphantom {{\Omega \left( \tau  \right)} \tau }} \right.
 \kern-\nulldelimiterspace} \tau }} \right].
\end{equation}

With $\Omega \left( \tau  \right) \to \infty $, the rest part of (\ref{12465523}) can converge to its stochastic average value. At this time, (\ref{12465523}) can be further determined as:
\begin{equation}\label{562}
AAoI = \mathop {\lim }\limits_{\tau  \to \infty } AAo{I_\tau } = \lambda \left( {\mathbb{E}\left[ {BT} \right] + \frac{1}{2}\mathbb{E}\left[ {{B^2}} \right]} \right).
\end{equation}

In LAWNs, the newly generated sensing data gets preemptive wireless channel access to ensure real-time transmission. The transfer procedure can be modeled by the first-come first-serve principle. Furthermore, according to \cite{yang2023stochastic}, the service rate of sensing data can be represented by the wireless channel's transmission rate of the sensing data. Therefore, the server (wireless channel) utilization can be defined as:
\begin{equation}\label{562332}
\rho  = {\lambda  \mathord{\left/
 {\vphantom {\lambda  {{\gamma _{sense}}}}} \right.
 \kern-\nulldelimiterspace} {{\gamma _{sense}}}}.
\end{equation}

As the low-altitude applications involve many time-sensitive targets, the generation and service processes of sensing data should be represented by appropriate queuing models, allowing the arrival and service rates to follow distinct probability distributions. Hence, three models are adopted here. Specifically, we first utilize M/M/1 model to describe unpredictable low-altitude airspace with irregular sensing data generation and varying service time due to the changing target signals, clutter, and interference. Second, we use D/M/1 model to  describe a time-division multiplexing sensing system with fixed time slots for target monitoring, but service time varies due to changing environmental conditions. Third, M/D/1 model is employed to describe a scenario where sensing data generation is random, but wireless channel condition is ideal enough, thereby sensing data has the deterministic service time.
Based on above analysis, we have the following theorems to derive the mathematical expressions of the AoI, thereby quantifying the freshness of the sensing data.

\textbf{Theorem 1:} In the case where both the generation and service time of sensing data  follow exponential distribution, the AoI of the sensing data can be expressed as:
\begin{equation}\label{145123}
AAo{I^{M/M/1}} = \frac{1}{{{\gamma _{sense}}}}\left( {1 + \frac{{{\gamma _{sense}}}}{\lambda } + \frac{{{\lambda ^2}}}{{\gamma _{sense}^2 - \lambda {\gamma _{sense}}}}} \right).
\end{equation}

\begin{IEEEproof}
Please refer to Appendix A for more details.
\end{IEEEproof}

\textbf{Theorem 2:} When the sensing data is deterministically generated, and the corresponding service time follows exponential distribution, the AoI of the sensing data can be expressed as:
\begin{equation}\label{942563256}
AAo{I^{D/M/1}} = \frac{1}{{{\gamma _{sense}}}}\left( {\frac{{{\gamma _{sense}}}}{{2\lambda }} + \frac{1}{{1 - \delta }}} \right),
\end{equation}
where $\delta  = {\mathcal{L}_B}\left( {{\gamma _{sense}}\left( {1 - \delta } \right)} \right)$, and $\mathcal{L}\left(  \cdot  \right)$ is the Laplace transform operator.

\begin{IEEEproof}
Please refer to Appendix B for more details.
\end{IEEEproof}

\textbf{Theorem 3:} In the case where the sensing data arrivals follow an exponential distribution and the service time is deterministic, then the AoI can be expressed as:
\begin{equation}\label{231223626}
AAo{I^{M/D/1}} = \lambda \int_0^\infty  {b\left( {{\xi _1}{\xi _2}} \right)} d{\Theta _U}\left( u \right) + Cons + \frac{1}{\lambda },
\end{equation}
where $Cons$ represents the fixed service time of sensing data, ${\xi _1}$, ${\xi _2}$, ${\chi _{Cons}}\left( b \right)$, and ${\Theta _U}\left( u \right)$ are defined as follows:
\begin{equation}\label{144552485}
{\xi _1} = \frac{{\left( {Cons} \right)\rho }}{{2\left( {1 - \rho } \right)}} + Cons - b - {\chi _{Cons}}\left( b \right),
\end{equation}
\begin{equation}\label{144552485234}
{\xi _2} = \int_0^{b - Cons} {\left( {u + Cons - b} \right)d{\Theta _U}\left( u \right)} ,
\end{equation}
\begin{equation}\label{1445524854556}
{\chi _{Cons}}\left( b \right) = \left\{ \begin{array}{l}
1,\quad b \ge Cons\\
0,\quad {\mathop{\rm otherwise}\nolimits}
\end{array} \right.,
\end{equation}
\begin{equation}\label{14455248726565}
{\Theta _U}\left( u \right) = \left( {1 - \rho } \right)\sum\nolimits_{i = 0}^{\left\lfloor {u/Cons} \right\rfloor } {{\rho ^i}} {\left( {i - \frac{u}{{Cons}}} \right)^i}{e^{\rho \left( {\frac{u}{{Cons}} - i} \right)}}.
\end{equation}

\begin{IEEEproof}
Please refer to Appendix C for more details.
\end{IEEEproof}

\section{ ISAC Performance Optimization for Time-Sensitive Targets under Attacks}\label{OP}

\subsection{Stackelberg Game-based Problem Formulation}

In the low-altitude airspace, the malicious drone launches the channel access attack and then adds noise to all the channels, deteriorating the ISAC performance of LAWNs. By using the channel estimation techniques, the legitimate drone and BS can detect the ISAC performance degradation. On this basis, LAWNs operators make corresponding adjustments to maintain the ISAC performance. Such a process matches the Stackelberg game with players including the malicious drone, the legitimate drone with an RIS, and the BS. Here, the malicious drone is the game leader, and the legitimate drone with RIS and BS are the vice and first follower, respectively. In this way, the leader first takes actions to maximize its utility. Then, the followers make corresponding changes to maximize their own utilities. Finally, after reaching the Stackelberg equilibrium, all players' utilities can converge, arriving at their own steady states. More concretely, two performance metrics are optimized in the game, including the average AoI of the sensing data defined in (\ref{145123}) to (\ref{231223626}) and the average communication SINR for all users, which is defined as follows:
\begin{equation}\label{521455}
{\mathop{\rm ASINR}\nolimits}  = \frac{1}{N}\sum\nolimits_{i = 1}^N {{\mathop{\rm SINR}\nolimits} _i^{com}}.
\end{equation}

\subsubsection{First Follower Sub-game}\
In the considered game, the BS acts as the first follower, it aims at minimizing the average AoI of sensing data and the cost of generating the sensing data, while maximizing the SINR of all the communication users. Therefore, the utility function of the BS can be defined as:
\begin{equation}\label{596211548696221}
{\varphi _{BS}} =  - {\zeta _1}AAoI + {\zeta _2}{\mathop{\rm ASINR}\nolimits}  - {\vartheta _{BS}}\lambda  ,
\end{equation}
where $\zeta _1$ and $\zeta _2$ are the coefficients to keep different performance metrics in the same order of magnitude, $\vartheta _{BS}$ represents the unit cost of generating sensing data with different rates. For the BS, its action is to adopt different rate strategies to produce the sensing data. Hence, the sub-game for the first follower can be expressed as:
\begin{equation}\label{5895614585}
\begin{aligned}
\text{SP}_{BS}: \quad & \underset{\lambda}{\text{max}} \; \varphi_{BS} \\
& \text{s.t.} \; 0 \le \lambda \le \gamma_{\text{radar}} \\
& \qquad \text{SINR}_i^{\text{com}} \ge \text{SINR}_{\text{thre}}
\end{aligned},
\end{equation}
where ${{\mathop{\rm SINR}\nolimits} _{thre}}$ is the threshold to ensure the communication link, $\lambda$ is the variable to be optimized by the BS, the constraint $0 \le \lambda $ can ensure that $\lambda$ is positive, and $\lambda  \le {\gamma _{sense}}$ indicates the data generation rate is smaller than the service rate, guaranteeing the queue converges to the steady state.

\subsubsection{Second Follower Sub-game}

The legitimate drone with RIS acts as the second follower in the established game. Similar to the BS, its optimization goal is to minimize the average AoI of sensing data and the power cost of the reflecting elements for amplifying the signal, while maximizing the SINR for all communication users. Therefore, the utility is defined as:
\begin{equation}\label{7854785}
{\varphi _{RIS}} =  - {\zeta _1}AAoI + {\zeta _2}{\mathop{\rm ASINR}\nolimits}  - {\vartheta _{RIS}}g,
\end{equation}
where ${\vartheta _{RIS}}$ is the unit cost of reflecting elements' power amplification. Besides, the legitimate drone with
RIS optimizes the reflecting elements' power amplification gain to maximize its utility. Thereby, the sub-game for the second follower is
\begin{equation}\label{4521486321}
\begin{aligned}
\text{SP}_{RIS}: \quad & \underset{g}{\text{max}} \; \varphi_{RIS} \\
& \text{s.t.} \; 0 \le g \le g_{\text{max}} \\
& \qquad \text{SINR}_i^{\text{com}} \ge \text{SINR}_{\text{thre}}
\end{aligned},
\end{equation}
where ${{\mathop{\rm SINR}\nolimits} _{thre}}$ is the SINR threshold to ensure the communication link, $g_{max}$ is the maximum power amplification gain of RIS's reflecting elements, which is constrained by the limited energy supply of the legitimate drone.

\subsubsection{Leader Sub-game}\

As for the  malicious drone conducting attack, it serves as the leader within the Stackelberg game. In contrast to the first follower and second follower, the optimization objective of the malicious drone is to maximize the average AoI of sensing data, while minimizing the SINR of all the communication users and the cost for adding noise to all available channels. On this basis, we let ${\varphi _{att}}$ be the utility function of the malicious drone, we have
\begin{equation}\label{44562}
{\varphi _{att}} = {\zeta _1}AAoI - {\zeta _2}{\mathop{\rm ASINR}\nolimits}  - {\vartheta _{att}}{\sigma _{att}},
\end{equation}
where $\vartheta _{att}$ represents the unit cost of adding noise to all the accessed channels. On this basis, the malicious drone can adjust the power of the added noise to maximize its utility, and therefore the leader sub-game can be formulated as:
\begin{equation}\label{96321456}
\begin{aligned}
\text{SP}_{\text{att}}: \quad & \underset{\sigma_{\text{att}}}{\text{max}} \; \varphi_{\text{att}} \\
& \text{s.t.} \; 0 \le \sigma_{\text{att}} \le \nu \sigma \\
& \qquad \text{SINR}_i^{\text{com}} < \text{SINR}_{\text{thre}}
\end{aligned}
\end{equation}
where ${\mathop{\rm SINR}\nolimits} _i^{com} < {{\mathop{\rm SINR}\nolimits} _{thre}}$ can ensure the effectiveness of the channel access attack. $\nu$ is a constant, ${\sigma _{att}} \le \ \nu\sigma$ is the constraint to ensure the attack's ability and covertness, ensuring that the added noise is similar to the environment noise to confuse the attack detection approaches.
\vspace{-0.3cm}
\subsection{Stackelberg Game Equilibrium Solution}
\subsubsection{Problem reformulation} Note that there is hierarchical conflicting relationship among leader and followers. Therefore, to further highlight the actions' order of all the players, based on the backward induction (BI) method~\cite{yang2022aoi}, the sub-problems in (\ref{5895614585}), (\ref{4521486321}), and (\ref{96321456}) can be reformulated as a three layer programming problem, which can be expressed as:
\begin{equation}\label{51155}
\begin{cases}
\begin{aligned}
& \underset{\sigma_{\text{att}}}{\text{max}} \; \varphi_{\text{att}}\left( {\lambda ,g,{\sigma _{\text{att}}}} \right) \\
& \text{s.t.} \;\quad 0 \le \sigma_{\text{att}} \le \sigma \\
& \text{SINR}_i^{\text{com}} < \text{SINR}_{\text{thre}} \\
& \text{Optimal solution:} \; \left( {{\lambda _{\text{opt}}},{g_{\text{opt}}},{\sigma _{\text{att}}}} \right)
\end{aligned} \\
\begin{cases}
\begin{aligned}
& \underset{g}{\text{max}} \; \varphi_{\text{RIS}}\left( {\lambda ,g,{\sigma _{\text{att}}}} \right) \\
& \text{s.t.} \;\quad 0 \le g \le {g_{\max }} \\
& \text{SINR}_i^{\text{com}} \ge \text{SINR}_{\text{thre}} \\
& \text{Optimal solution:} \; \left( {{\lambda _{\text{opt}}},g,{\sigma _{\text{att}}}} \right)
\end{aligned} \\
\begin{cases}
\begin{aligned}
& \underset{\lambda}{\text{max}} \; \varphi_{\text{BS}}\left( {\lambda ,g,{\sigma _{\text{att}}}} \right) \\
& \text{s.t.} \;\quad 0 \le \lambda \le {\gamma _{\text{radar}}} \\
& \text{SINR}_i^{\text{com}} \ge \text{SINR}_{\text{thre}}
\end{aligned}
\end{cases}
\end{cases}
\end{cases},
\end{equation}
where ${{\lambda _{opt}}}$ and ${{g_{opt}}}$ represent optimized values for variable $\lambda$ and $g$, respectively. Following the BI approach, three sub-problems are solved from bottom to top, which can provide the optimal generation rate of sensing data, optimal power of reflecting elements' amplification, as well as the optimal values of the added noise in order.

\subsubsection{GSSPI-based Sub-problems Solution}

To solve the problem in (\ref{51155}), the first step is to handle the three sub-problems. Due to the lack of closed-form solutions for the objective functions, Golden section search and parabolic interpolation (GSSPI) methods-based approach~\cite{yang2024can} is improved here to derive the optimized solution of each sub-problem. Generally, GSSPI is a derivative-free optimization technique that does not rely on gradient information, making it well-suited for sub-problems that are noisy or non-differentiable. GSSPI can refine the search interval using golden section search's guaranteed convergence, and accelerates the process with parabolic interpolation's quadratic convergence, effectively handling the constraints. Specifically, GSSPI first narrows the search interval through the golden section method, whose convergence is guaranteed. Then, it accelerates the optimization with parabolic interpolation, which achieves quadratic convergence while satisfying the imposed constraints.

The GSSPI is designed for minimization problems, and hence we define $\Psi  = \left\{ { - {\varphi _{BS}}, - {\varphi _{RIS}}, - {\varphi _{att}}} \right\}$ as the objective function to be minimized and $\Lambda  = \left\{ {\lambda ,g,{\sigma _{att}}} \right\}$ as the variable to be optimized. Besides, let ${\Lambda _{min }}$, ${\Lambda _{max }}$ and ${\Lambda _{opt }}$ represent the lower bound, upper bound, and the optimized variable, respectively. On this basis, the GSSPI-based algorithm for sub-problem is detailed in Algorithm 1. Specifically, Step 1--Step 2 first initialize the loop settings. Then, Step 3 determines the global optimal domain, reducing the search range of the solution~\cite{forsythe1977computer,brent2013algorithms}. Next, Step 4 determines the solution ${\Lambda _{opt}(Iter)}$, and Step 5 clarifies the loop ending conditions. At last, Step 6 determines the final value of the target function.

\begin{algorithm}[t]
	\renewcommand{\algorithmicrequire}{\textbf{Input:}}
	\renewcommand{\algorithmicensure}{\textbf{Output:}}
	\caption{GSSPI-based algorithm for sub-problem solving }
	\label{algor1}
	\begin{algorithmic}[1]
		\REQUIRE $\Psi $: the target function to be minimized; $\Lambda $: the variable to be optimized; $\left[ {{\Lambda _{min }},{\Lambda _{max }}} \right]$: the feasible domain of the variable; $Iter=1$: the looping iteration index; $Iter_{max}$: the maximum iteration index; ${\Lambda _{1}}$: the initial variable value; $\varsigma $: the error tolerance threshold.
		\ENSURE ${\Lambda _{opt}}$: the optimized variable value; $\Psi \left( {{\Lambda _{opt}}} \right)$: the optimal value of the target function.
        \STATE \textbf{Repeat}
        \STATE \quad $Iter = Iter +1 $
        \STATE \quad Based on the Golden section searching, determine the global optimal domain $\left[ {{\Lambda _{opt-min }(Iter)},{\Lambda _{opt-max }(Iter)}} \right]$ for variable $\Lambda$
        \STATE \quad Based on the parabolic interpolation, determine the solution ${\Lambda _{opt}(Iter)}$ within $\left[ {{\Lambda _{opt-min }(Iter)},{\Lambda _{opt-max }(Iter)}} \right]$
        \STATE  \textbf{Until} $\left| {{\Lambda _{opt}}\left( {Iter} \right) - {\Lambda _{opt}}\left( {Iter - 1} \right)} \right| \le \varsigma $ or $Iter > Ite{r_{max }}$
        \STATE Calculate $\Psi \left( {{\Lambda _{opt}}\left( {Iter} \right)} \right)$ based on ${{\Lambda _{opt}}\left( {Iter} \right)}$
\end{algorithmic}
\end{algorithm}


\subsubsection{Backward Induction Stackelberg Equilibrium Solution}

To simultaneously maximize all the players' utilities, we proposed a BI-based algorithm to calculate the Stackelberg equilibrium, as shown in Algorithm 2. Specifically, Step 1 to Step 3 first initialize the parameters of all players' actions and utilities. Step 4 to Step 6 determine the looping rule and iteration index. Next, through Step 7 to Step 13, the first follower sub-game can be solved. Similarly, the second follower sub-game and leader sub-game can also be solved through Step 14 to Step 20, and Step 21 to Step 27, respectively. Finally, Step 28 to Step 30 end the loop, and determine the output results as the Stackelberg equilibrium solutions and utilities.

\begin{algorithm}[t]
	\renewcommand{\algorithmicrequire}{\textbf{Input:}}
	\renewcommand{\algorithmicensure}{\textbf{Output:}}
	\caption{Stackelberg equilibrium calculation}
	\label{algor2}
	\begin{algorithmic}[1]
		\REQUIRE ${\varphi _{BS}}$, $\lambda $, $\left[ {0,{\gamma _{sense}}} \right]$, ${\varphi _{RIS}}$, $g$, $\left[ {0,{g_{\max }}} \right]$, ${\varphi _{att}}$, ${\sigma _{att}}$, $\left[ {0,\sigma } \right]$, $Iter'=1$, and $Iter'_{max}$.        
        \ENSURE ${{\lambda _{opt}}}$, ${{g_{opt}}}$, ${\sigma _{att\_opt}}$, ${\varphi _{BS}}_{\_opt}$, ${\varphi _{RIS}}_{\_opt}$, and ${\varphi _{att}}_{\_opt}$.
        \STATE \textbf{//Stage 1: Parameters initialization}
        \STATE  Initialize parameters of all the players' actions: ${\lambda _{opt}}\left( 1 \right)$, ${g _{opt}}\left( 1 \right)$, and ${\sigma _{att\_opt}}\left( 1 \right)$
        \STATE Initialize parameters of all the players' utilities with (\ref{596211548696221}), (\ref{7854785}), and (\ref{44562}): ${\varphi _{BS}}_{\_opt}\left( 1 \right)$, ${\varphi _{RIS}}_{\_opt}\left( 1 \right)$, and ${\varphi _{att}}_{\_opt}\left( 1 \right)$
        \STATE \textbf{//Stage 2: Repeat iterations}
        \STATE \textbf{While} $Iter'  <  Ite{r'_{\max }}$ \textbf{do}
        \STATE \quad $Iter' = Iter' +1 $
        \STATE \textbf{//Stage 2-1: First follower sub-game solving}
        \STATE \quad Derive ${\lambda _{opt}}\left( {Iter} \right)$ with \textbf{Algorithm 1}
        \STATE \quad Based on (\ref{596211548696221}), calculate the current utility ${\varphi _{BS}}_{\_opt}\left( {Iter} \right)$ with ${\lambda _{opt}}\left( {Iter} \right)$, ${g_{opt}}\left( {Iter - 1} \right)$ and ${\sigma _{att\_opt}}\left( {Iter - 1} \right)$
        \STATE \quad \textbf{If} ${\varphi _{BS}}_{\_opt}\left( {Iter} \right) < {\varphi _{BS}}_{\_opt}\left( {Iter - 1} \right)$ holds on
        \STATE \quad \quad Let ${\lambda _{opt}}\left( {Iter} \right) = {\lambda _{opt}}\left( {Iter - 1} \right)$
        \STATE \quad \quad Let ${\varphi _{BS}}_{\_opt}\left( {Iter} \right) = {\varphi _{BS}}_{\_opt}\left( {Iter - 1} \right)$
        \STATE \quad \textbf{End}

        \STATE \textbf{//Stage 2-2: Second follower sub-game solving}
        \STATE \quad Derive ${g _{opt}}\left( {Iter} \right)$ with \textbf{Algorithm 1}
        \STATE \quad  Based on (\ref{7854785}), calculate the current utility ${\varphi _{RIS}}_{\_opt}\left( {Iter} \right)$ with ${\lambda _{opt}}\left( {Iter} \right)$, ${g_{opt}}\left( {Iter} \right)$ and ${\sigma _{att\_opt}}\left( {Iter - 1} \right)$
        \STATE \quad \textbf{If} ${\varphi _{RIS}}_{\_opt}\left( {Iter} \right) < {\varphi _{RIS}}_{\_opt}\left( {Iter - 1} \right)$ holds on
        \STATE \quad \quad Let ${g _{opt}}\left( {Iter} \right) = {g _{opt}}\left( {Iter - 1} \right)$
        \STATE \quad \quad Let ${\varphi _{RIS}}_{\_opt}\left( {Iter} \right) = {\varphi _{RIS}}_{\_opt}\left( {Iter - 1} \right)$
        \STATE \quad \textbf{End}

        \STATE \textbf{//Stage 2-3: Leader sub-game solving}
        \STATE \quad Derive ${\sigma _{att\_opt}}\left( {Iter} \right)$ with \textbf{Algorithm 1}
        \STATE \quad  Based on (\ref{44562}), calculate the current utility ${\varphi _{att}}_{\_opt}\left( {Iter} \right)$ with ${\lambda _{opt}}\left( {Iter} \right)$, ${g_{opt}}\left( {Iter} \right)$ and ${\sigma _{att\_opt}}\left( {Iter } \right)$
        \STATE \quad \textbf{If} ${\varphi _{att}}_{\_opt}\left( {Iter} \right) < {\varphi _{att}}_{\_opt}\left( {Iter - 1} \right)$ holds on
        \STATE \quad \quad Let ${\sigma _{att\_opt}}\left( {Iter} \right) = {\sigma _{att\_opt}}\left( {Iter - 1} \right)$
        \STATE \quad \quad Let ${\varphi _{att}}_{\_opt}\left( {Iter} \right) = {\varphi _{att}}_{\_opt}\left( {Iter - 1} \right)$
        \STATE \quad \textbf{End}
        \STATE \textbf{//Stage 3: End iterations}
        \STATE \textbf{End}
        \STATE Determine ${{\lambda _{opt}}}$, ${{g_{opt}}}$, ${\sigma _{att\_opt}}$, ${\varphi _{BS}}_{\_opt}$, ${\varphi _{RIS}}_{\_opt}$, ${\varphi _{att}}_{\_opt}$ as the Stackerberg equilibrium solutions and utilities.
\end{algorithmic}
\end{algorithm}

\subsubsection{Complexity and Convergence Analysis}
According to \cite{forsythe1977computer} and \cite{brent2013algorithms}, the computational complexity of Algorithm~\ref{algor1} is bounded by $o\left\{ {\log \left[ {{{\left( {{\Lambda _{\max }} - {\Lambda _{\min }}} \right)} \mathord{\left/
 {\vphantom {{\left( {{\Lambda _{\max }} - {\Lambda _{\min }}} \right)} \varsigma }} \right.
 \kern-\nulldelimiterspace} \varsigma }} \right]} \right\}$ and $o\left( {{{\left( {Ite{r_{\max }}} \right)}^2}} \right)$. In addition, the convergence of Algorithm~\ref{algor1} is primarily determined by the search step and tolerance threshold. Because the search range is bounded, Algorithm \ref{algor1} is guaranteed to converge in only a few iterations. At each step it reduces the bracketing interval monotonically by a factor of about 0.618 without requiring derivative information, so the optimum remains inside an interval that shrinks at a linear rate. Once this interval becomes sufficiently small, the GSSPI procedure typically attains the optimum quickly because the local behavior of many functions near a minimum is well approximated by a parabola. For the proposed BI-based Algorithm~\ref{algor2} for Stackelberg equilibrium calculating , according to \cite{yang2022aoi}, its complexity mainly depends on Stage 2-1 to Stage 2-3. Based on Section V-B-3), one can see that the complexity of calculating the player's utility is $o\left( \varpi  \right)$, where $\varpi$ is a constant. Hence, the complexity of solving a single player's sub-game one time is bounded by $o\left\{ {\log \left[ {{{\left( {{\Lambda _{\max }} - {\Lambda _{\min }}} \right)} \mathord{\left/
 {\vphantom {{\left( {{\Lambda _{\max }} - {\Lambda _{\min }}} \right)} \varsigma }} \right.
 \kern-\nulldelimiterspace} \varsigma }} \right]} \right\} + o\left( \varpi  \right)$ and $o\left\{ {{{\left( {Ite{r_{\max }}} \right)}^2}} \right\} + o\left( \varpi  \right)$. In this regard, the total complexity of Algorithm~\ref{algor2} is within $o\left( 1 \right) + o\left\{ {3Ite{{r'}_{\max }}\left[ {o\left( {\log \left( {{{\left( {{\Lambda _{\max }} - {\Lambda _{\min }}} \right)} \mathord{\left/
 {\vphantom {{\left( {{\Lambda _{\max }} - {\Lambda _{\min }}} \right)} \varsigma }} \right.
 \kern-\nulldelimiterspace} \varsigma }} \right)} \right) + o\left( \varpi  \right) + o\left( 1 \right)} \right]} \right\}$ and $o\left( 1 \right) + o\left\{ {3Ite{{r'}_{\max }}\left[ {o\left( {{{\left( {Ite{r_{\max }}} \right)}^2}} \right) + o\left( \varpi  \right) + o\left( 1 \right)} \right]} \right\}$. In addition, the convergence of Algorithm~\ref{algor2} follows from that of Algorithm~\ref{algor1}. Section V-B-3) shows that Algorithm~\ref{algor1} converges in a finite number of iterations. Because Algorithm~\ref{algor2} invokes Algorithm~\ref{algor1} only a finite number of times, it also reaches a stable solution within limited iterations.

\subsection{Stackelberg Equilibrium Analysis}

For the formulated Stackelberg game in (\ref{51155}), its Stackelberg equilibrium can be determined as follows:
\begin{equation}\label{14521485254896}
\left\{ \begin{array}{l}
{\varphi _{BS}}\left( {{\lambda _{opt}},{g_{opt}},{\sigma _{att\_opt}}} \right) \ge {\varphi _{BS}}\left( {\lambda ,{g_{opt}},{\sigma _{att\_opt}}} \right)\\
{\varphi _{RIS}}\left( {{\lambda _{opt}},{g_{opt}},{\sigma _{att\_opt}}} \right) \ge {\varphi _{BS}}\left( {\lambda ,g,{\sigma _{att\_opt}}} \right)\\
{\varphi _{att}}\left( {{\lambda _{opt}},{g_{opt}},{\sigma _{att\_opt}}} \right) \ge {\varphi _{BS}}\left( {\lambda ,{g_{opt}},{\sigma _{att}}} \right)
\end{array} \right..
\end{equation}
\textcolor{black}{For the above-defined Stackelberg equilibrium, we have the following theorem to ensure its existence and uniqueness.}

\textbf{Theorem 4:} For the formulated Stackelberg game in (\ref{51155}), its Stackelberg equilibrium always exists and is unique.
\begin{IEEEproof}
According to \cite{aubin2007mathematical}, by combining the followers' best response mapping with the leader's strategy space, a mapping from the strategy space to the leader itself can be created. If this mapping satisfies the conditions of a fixed-point theorem \cite{aubin2007mathematical}, then a fixed point exists, which corresponds to a Stackelberg equilibrium. Specifically, the followers' best response mapping can be represented as follows:
\begin{equation}\label{bdd1}
{R_{F\_RIS}}\left( g \right) = \arg {\max _{g \in \left[ {0,{g_{\max }}} \right]}}{\varphi _{RIS}}\left( {g,\lambda ,{\sigma _{att}}} \right),    
\end{equation}
\begin{equation}\label{bdd2}
{R_{F\_BS}}\left( \lambda  \right) = \arg {\max _{\lambda  \in \left[ {0,{\gamma _{sense}}} \right]}}{\varphi _{BS}}\left( {\lambda ,g,{\sigma _{att}}} \right).   
\end{equation}

On this basis, the leader can derive its strategy with: 
\begin{equation}\label{bdd3}
\sigma {*_{att}} = \arg {\max _{{\sigma _{att}} \in \left[ {0,\nu \sigma } \right]}}{\varphi _{att}}\left( {\lambda ,g,{\sigma _{att}}} \right).  
\end{equation} 

Then, we can further construct a mapping $\Xi \left( {{\sigma _{att}}} \right)$ from leader's strategy space to itself, mathematically,
\begin{equation}\label{bdd4}
\begin{array}{l}
\Xi \left( {{\sigma _{att}}} \right) = \\
\arg {\max _{\sigma {'_{att}} \in \left[ {0,\nu \sigma } \right]}}{\varphi _{att}}\left( {\sigma {'_{att}},{R_{F\_RIS}}\left( {{\sigma _{att}}} \right),{R_{F\_BS}}\left( {{\sigma _{att}}} \right)} \right)
\end{array}.  
\end{equation} 

In this regard, based on the fixed-point theorem \cite{aubin2007mathematical}, if the leader's strategy space is a non-empty, compact, and convex set in Euclidean space, and the constructed mapping is a continuous mapping, then the constructed mapping has a fixed point, corresponding to a Stackelberg equilibrium. Since the leader's strategy space ${\left[ {0,\nu \sigma } \right]}$ is non-empty, compact, and convex, and the constructed mapping $\Xi \left( {{\sigma _{att}}} \right)$ is a continuous single-value function, satisfying the continuous mapping needs. Finally, the fixed point exists and is unique, and Theorem 4 is proved.

\end{IEEEproof}

\section{Evaluations and Analysis}\label{EA}

This evaluates the proposed algorithm under various configurations. First, we evaluate the convergence property of the proposed algorithm from the perspectives of utility, optimized variables, and performance metrics. Next, we compare the proposed method with the three existing baselines in terms of utility. The baselines contain lightweight solutions and computationally intensive approaches. The former includes the average and random resource allocation schemes, and the latter refers to the Genetic Algorithm (GA)-based optimization scheme. Finally, we further compare the derived Stackelberg equilibrium solution with the Nash equilibrium~\cite{yang2022aoi} solution to demonstrate the superiority of Stackelberg equilibrium.
\vspace{-0.3cm}
\subsection{Parameter Setting}
In the low-altitude air space, with the help of legitimate drone equipped with RIS, one BS with multi-antenna is used to simultaneously communicate with two signal-antenna users and sense one target. 
The simulation parameters are presented in Table \ref{parameters}, and the communication SINR threshold is set as 5 dB. Furthermore, average allocation, random allocation, and GA-based optimization are introduced to make comparison with the proposed method. Specifically, the average scheme allocates resources, such as noise power of attacker and sensing data generation rate, with the average value across the feasible domain~\cite{chang2021resource}. This ensures fairness by equally allocating resources to all players regardless of channel quality or their individual needs and provides a neutral baseline for comparison. In the random allocation scheme~\cite{birhanie2020stochastic}, each resource value is independently sampled from a uniform distribution within its feasible range and assigned to a player. Since this process is purely random and does not consider the current system state, it serves as a baseline for unstructured and non-intelligent resource management. In the GA-based optimization scheme~\cite{ma2024optimizing}, a population of candidate solutions is iteratively evolved using selection, crossover, and mutation operations to obtain the optimal solution. This approach captures the complex relationship between resource allocation and system performance, enabling convergence toward superior solutions with a higher computational cost.
\vspace{-0.3cm}

\begin{table}[]
\centering
\caption{Simulation settings.}
\label{parameters}
\resizebox{0.48\textwidth}{!}{
\begin{tabular}{c c c c}
\toprule
\textbf{Symbol} & \textbf{Value} & \textbf{Symbol} & \textbf{Value} \\
\midrule
$M$ & 4 & $\sigma^2$ & -174 dBm/Hz \\
$N$ & 2 & $P$ & 16 \\
$(x_{BS}, y_{BS}, z_{BS})$ & $[0, 0, 1.5]$ & $(x_{RIS}, y_{RIS}, z_{RIS})$ & $[5, 10, 1.5]$ \\
$(x_{user}, y_{user}, z_{user})$ & $[30, 10, 1.5]$ & $(x_{tar}, y_{tar}, z_{tar})$ & $[0, 60, 1.5]$ \\
$r_{user}$ & 10 & $\varepsilon $ & 0.1 \\
$\eta$ & $10^7$ Hz & $P_{trans}$ & 0.9 \\
$\sigma^2_{y}$ & -174 dBm/Hz & $\sigma^2_{r}$ & -174 dBm/Hz \\
$g_{max}$ & 1 & $Iter_{max}$ & 25 \\
\bottomrule
\end{tabular}
}
\vspace{-0.3cm}
\end{table}

\subsection{Convergence Analysis}

\subsubsection{Convergence of the Utility Functions}
\begin{figure*}[!htb]
\centering
\subfigure[]
{
    \label{tu1}
    \includegraphics[width=0.63\columnwidth]{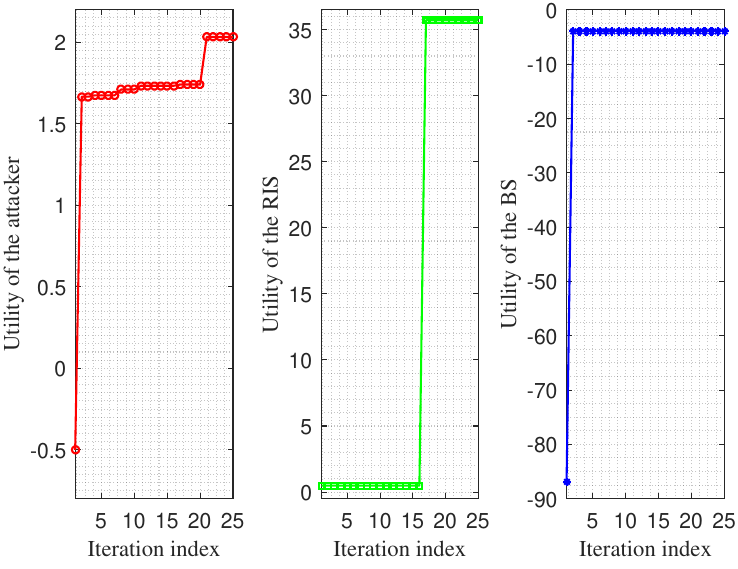}
}
\qquad
\hspace{-0.3in}
\subfigure[]
{
   \label{tu2}
    \includegraphics[width=0.63\columnwidth]{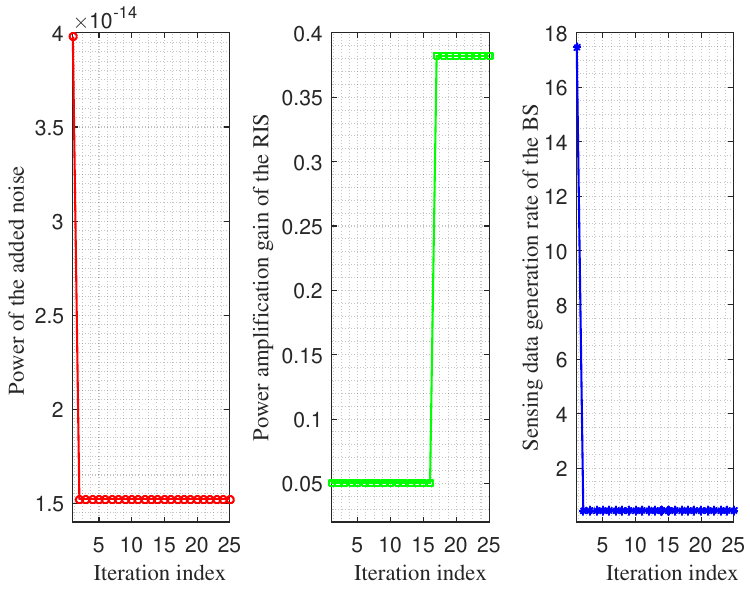}
}
\qquad
\hspace{-0.3in}
\subfigure[]
{
   \label{tu3}
    \includegraphics[width=0.6\columnwidth]{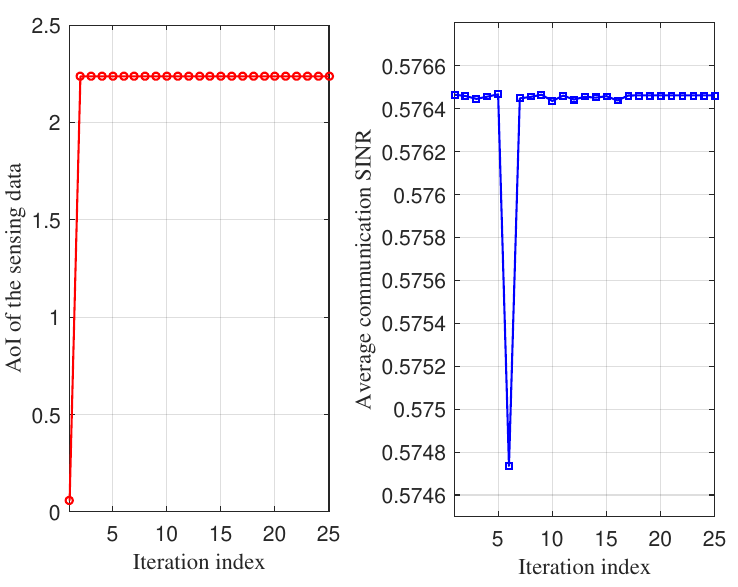}
}
\quad
\hspace{-0.3in}
\caption{Convergence behaviors evaluation of utility functions, performance metrics, and optimized variables. (a) Convergence behavior of the utility functions. (b) Convergence behavior of the optimized variables. (c) Convergence behavior of the performance metrics.}
\label{tu123}
\end{figure*}

\
Fig. \ref{tu1} is the convergence evaluation of utility functions. As shown in Fig. \ref{tu1},  we can see that as the number of iterations grows, utilities of the attacker, legitimate drone with RIS, and BS all can reach the stable values, confirming the convergence property of the proposed scheme. Moreover, each player’s utility rises throughout the iterations. This is because the established Stackelberg game is not cooperative and each participant is selfish and rational, aiming to maximize its own utility during the iterative optimization process.

\subsubsection{Convergence of Variables}\
Fig. \ref{tu2} evaluates the convergence of the optimized variables, including the sensing data generation rate of the BS, power amplification gain of the RIS, and the power of the added noise from attacker. Specifically, following the backward induction order, the BS, as the first follower, optimizes its sensing data generation rate and converges to 0.447. Then, the drone with RIS, acting as the second follower, optimizes the power amplification gain and reaches stability at 0.382. Finally, the attacker, as the game leader, optimizes the added noise power and reaches a stable value of $1.52 \times {10^{ - 14}}$. Therefore, within the finite iterations, all above-mentioned variables can converge.

\subsubsection{Convergence of the Performance Metrics} Fig. \ref{tu3} presents the convergence evaluation of performance metrics. As can be seen from Fig. \ref{tu3}, within the finite iteration steps, by optimizing the sensing data generation rate of the BS, power amplification gain of the RIS, and added noise power of the attacker in order, the formulated game can reach the Stackelberg equilibrium. At this time, the AoI of the sensing data and average communication SINR converge to 2.236 and 0.576, respectively.
\vspace{-0.3cm}
\subsection{Effectiveness Analysis}
\subsubsection{Performance Versus the Number of RIS Elements} 
\begin{figure*}[!htb]

\centering
\subfigure[]
{
    \label{tu4}
    \includegraphics[width=0.55\columnwidth]{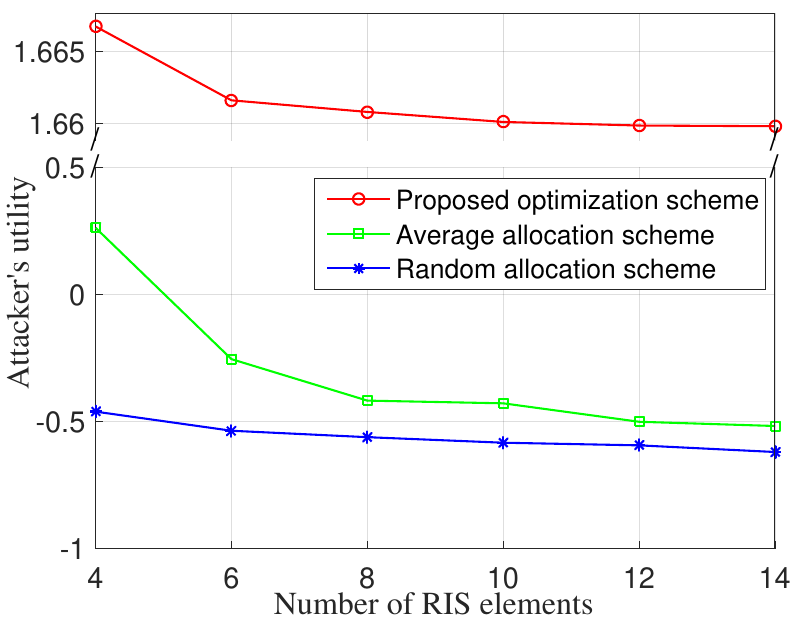}
}
\qquad
\hspace{-0.3in}
\subfigure[]
{
   \label{tu5}
    \includegraphics[width=0.55\columnwidth]{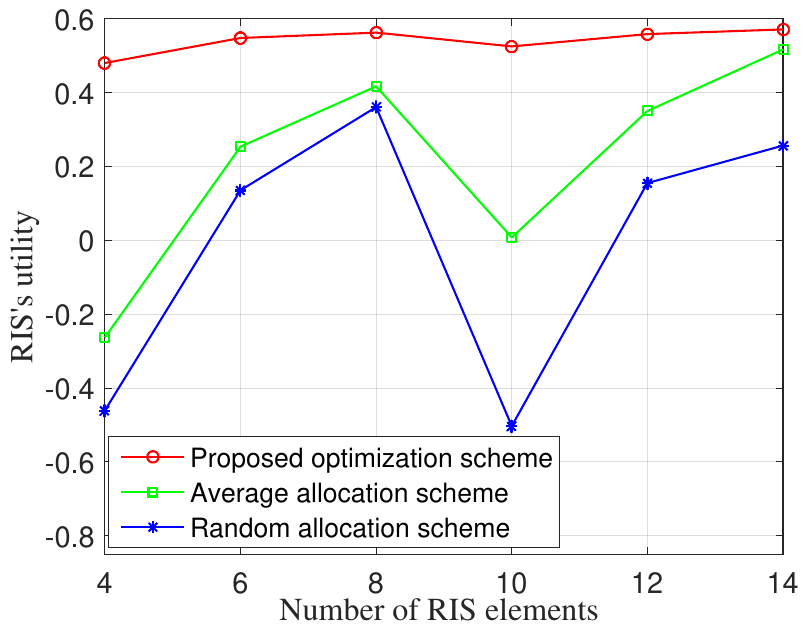}
}
\qquad
\hspace{-0.3in}
\subfigure[]
{
   \label{tu6}
    \includegraphics[width=0.55\columnwidth]{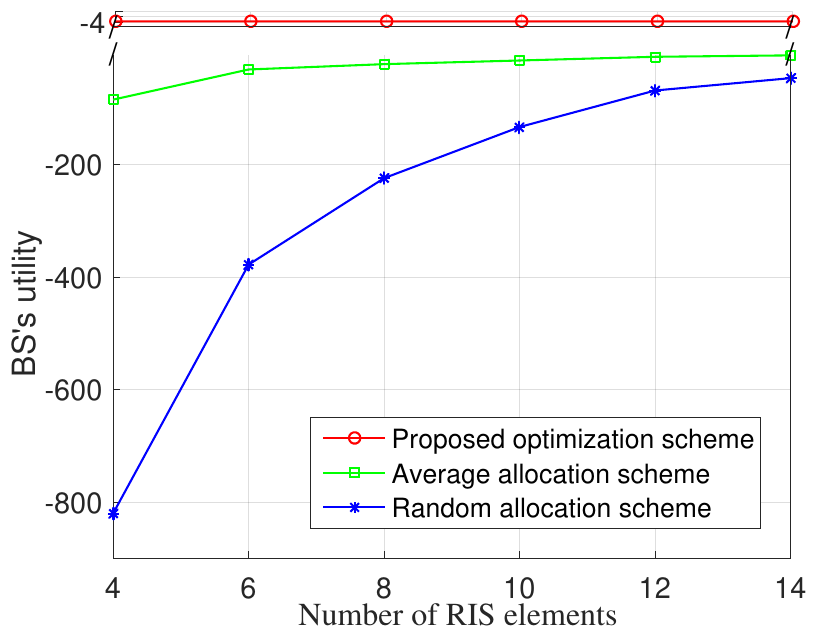}
}
\quad
\hspace{-0.3in}
\caption{Utility functions comparison of the attacker, RIS, and BS with different number of RIS elements. (a) The attacker's utility with different RIS element number. (b) The RIS's utility with different RIS element number. (c) The BS's utility with different RIS element number.}
\label{tu456}
\vspace{-0.3cm}
\end{figure*}

Fig. \ref{tu456} is the utility comparison of the attacker, RIS, and BS with different number of RIS elements. As shown in Fig.~\ref{tu456}, the proposed scheme achieves utility of 1.659, 0.571, and -3.895 for the attacker, RIS, and BS, respectively, outperforming those of the average-allocation scheme (-0.517, 0.517, and -6.28) and the random-allocation scheme (-0.62, 0.256, and -46.718). This demonstrates the effectiveness of the proposed approach. Moreover, Fig.~\ref{tu4} shows that the attacker’s utility falls as the number of RIS elements grows. The reason is that more elements improves the SINR, thereby reducing the attacker's utility. Furthermore, from Fig. \ref{tu5}, we can see that the RIS's utility first increases, then decreases, and increases again. This is because, with 8 or 14 RIS elements, the payoff is larger than the additional cost, whereas with 10 elements the payoff is lower than the cost, reducing overall utility. For the BS, Fig.~\ref{tu6} shows that its utility rises as the number of RIS elements increases. This can be interpreted by that more elements enhance signal strength, which raises the average SINR. The higher SINR increases channel capacity, allowing sensing data to be transmitted without delay, thereby reducing the AoI. Hence, as reflected in (\ref{596211548696221}), the BS utility grows.

\subsubsection{Performance Versus the Number of Transmitting Antennas} 
\begin{figure*}[!htb]
\centering
\subfigure[ ]
{
    \label{tu7}
    \includegraphics[width=0.55\columnwidth]{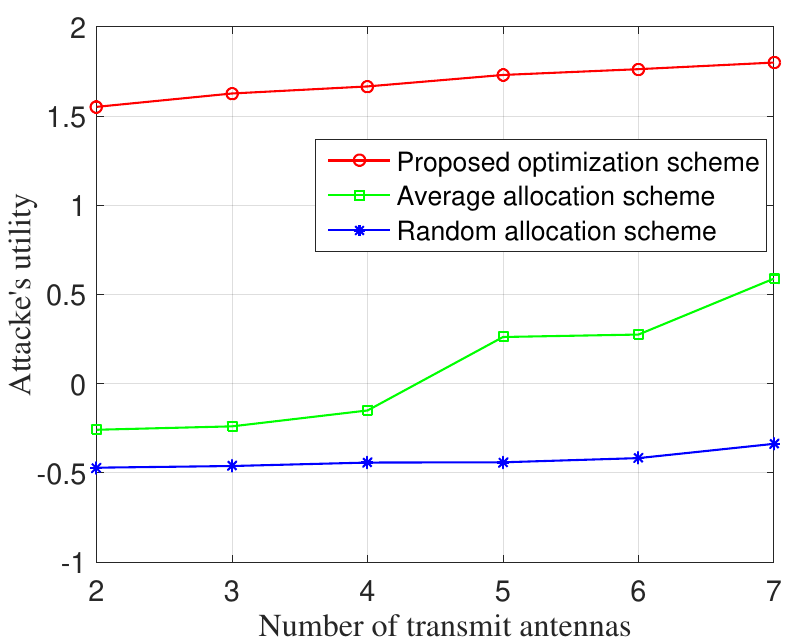}
}
\qquad
\hspace{-0.3in}
\subfigure[ ]
{
   \label{tu8}
    \includegraphics[width=0.55\columnwidth]{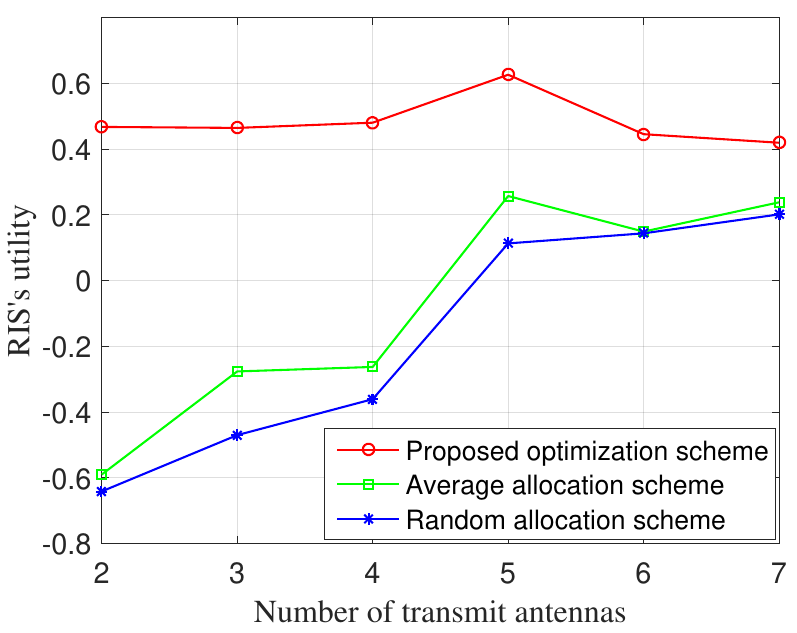}
}
\qquad
\hspace{-0.3in}
\subfigure[ ]
{
   \label{tu9}
    \includegraphics[width=0.56\columnwidth]{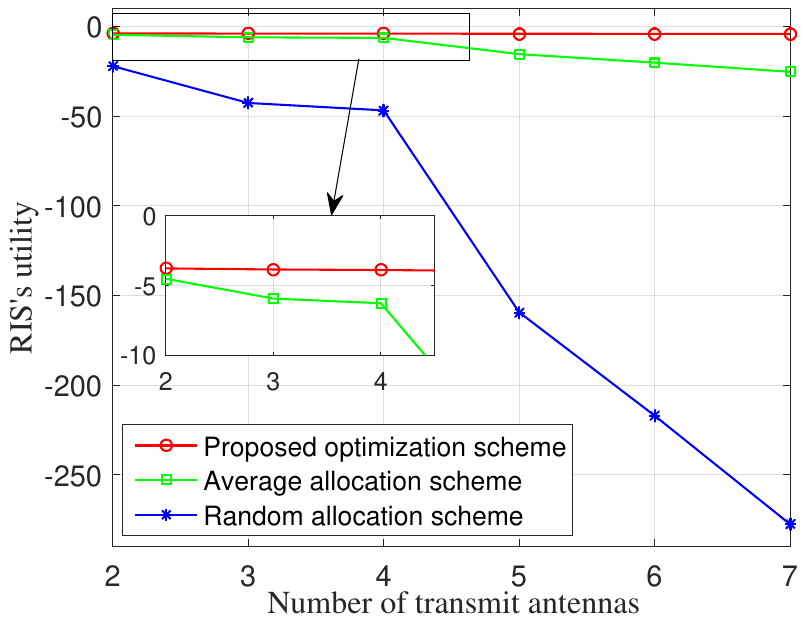}
}
\quad
\hspace{-0.3in}
\caption{Utility functions comparison of the attacker, RIS, and BS with different number of transmit antennas. (a) The attacker's utility with different transmit antenna number. (b) The RIS's utility with different transmit antenna number. (c) The BS's utility with different transmit antenna number.}
\label{tu789}
\end{figure*}

Fig.~\ref{tu789} presents the utility comparison of the attacker, RIS, and BS with different number of BS's transmitting antennas. As can be seen, across all cases, all players can obtain higher utilities by using the proposed optimization scheme compared to the average and random allocation schemes. Besides, different players obtain different benefits when the number of transmit antennas increases. Specifically, Fig.~\ref{tu7} shows that the attacker’s utility rises with more transmit antennas, because a multi-antenna system enlarges the signal’s spatial dimension, broadening the effective range of the attack. Fig.~\ref{tu8} shows that the RIS's utility fluctuates with the increasing number of transmit antenna. With 2, 3, 6, and 7 antennas at the BS, the RIS’s utility gain is smaller than that of the attacker, hence its utility declines. In contrast, with 4 or 5 antennas, the multi-antenna benefit outweighs the attacker’s advantage, thereby raising the RIS utility. Moreover, in Fig.~\ref{tu9}, we can see that, unlike attacker and RIS, the BS's utility decreases as the number of transmit antennas increases. This is because the multi-antenna system can increase the BS's transmitting capacity. To maintain the AoI performance, the BS needs to increase the sensing data generation rate, which increases cost and therefore reduces the BS’s overall utility.

\subsubsection{Performance Versus Different Self-interference Coefficient} 
\begin{figure*}[!htb]
\centering
\subfigure[ ]
{
    \label{tu10}
    \includegraphics[width=0.58\columnwidth]{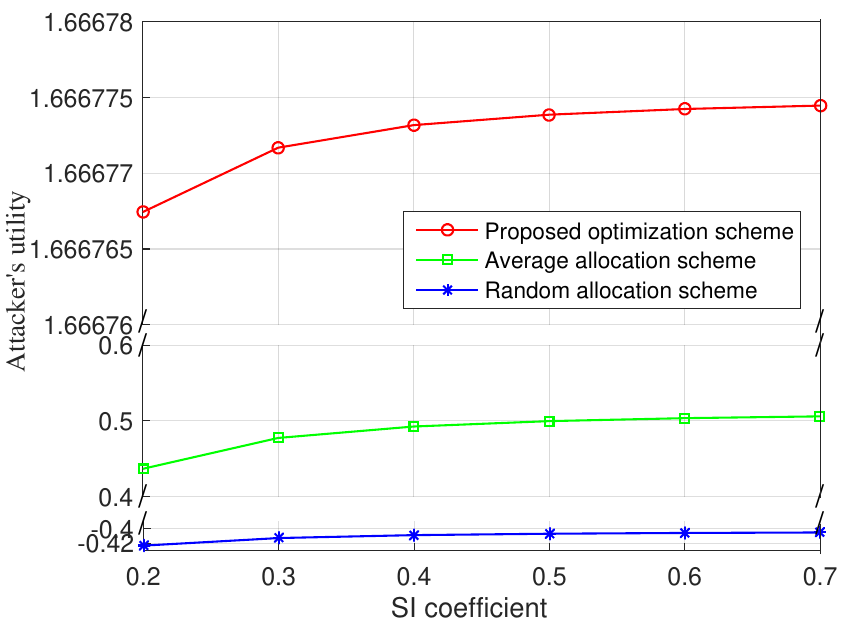}
}
\qquad
\hspace{-0.3in}
\subfigure[ ]
{
   \label{tu11}
    \includegraphics[width=0.55\columnwidth]{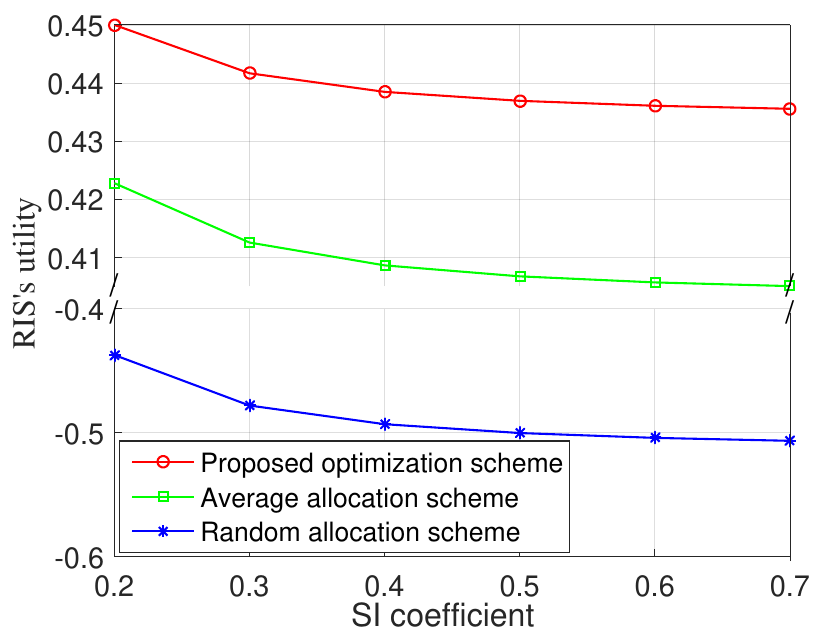}
}
\qquad
\hspace{-0.3in}
\subfigure[ ]
{
   \label{tu12}
    \includegraphics[width=0.58\columnwidth]{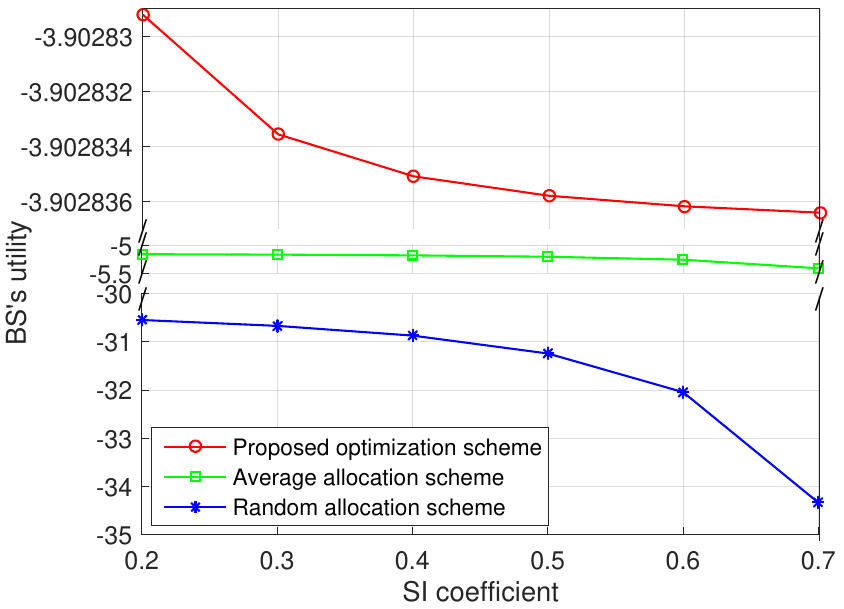}
}
\quad
\hspace{-0.3in}
\caption{Utility functions comparison of the attacker, RIS, and BS with different self-interference coefficient values. (a) The attacker's utility with different self-interference coefficient values. (b) The RIS's utility with different self-interference coefficient values. (c) The BS's utility with different self-interference coefficient values.}
\vspace{-0.3cm}
\label{tu101112}
\end{figure*}

Fig.~\ref{tu101112} compares the utility the attacker, RIS, and BS with different self-interference coefficient values. Overall, across different interference levels, the proposed scheme consistently outperforms all baselines, further demonstrating its superiority. Concretely, Fig.~\ref{tu10} shows that the attacker’s utility rises as the self-interference coefficient increases. The reason is that according to (\ref{11531321}), a higher self-interference coefficient raises the interference-plus-noise level, which lowers both AoI and average SINR, thereby enhancing attacker’s utility. For the same reason, Figs.~\ref{tu11} and \ref{tu12} reveal that the utilities of the RIS and the BS move in the opposite direction.

\subsubsection{Performance Comparison}
\begin{figure*}[!htb]
\centering
\subfigure[ ]
{
    \label{tuadd1}
    \includegraphics[width=0.61\columnwidth]{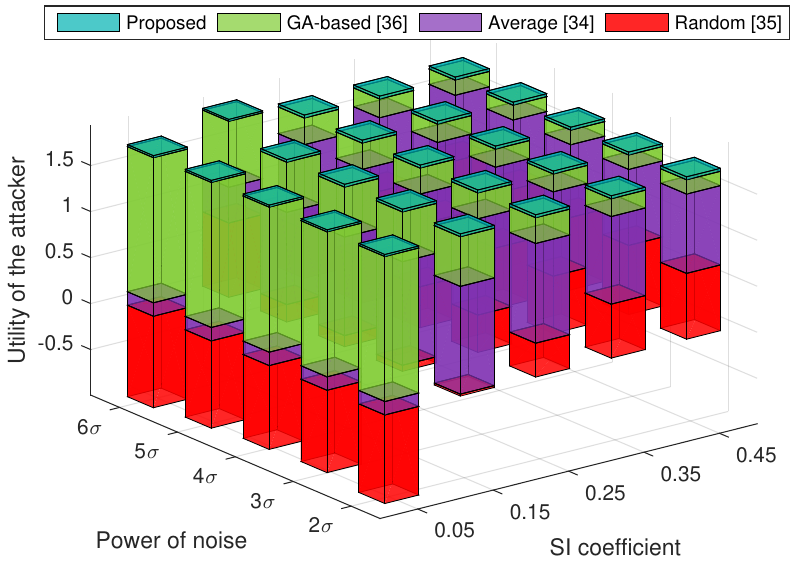}
}
\qquad
\hspace{-0.3in}
\subfigure[ ]
{
   \label{tuadd2}
    \includegraphics[width=0.6\columnwidth]{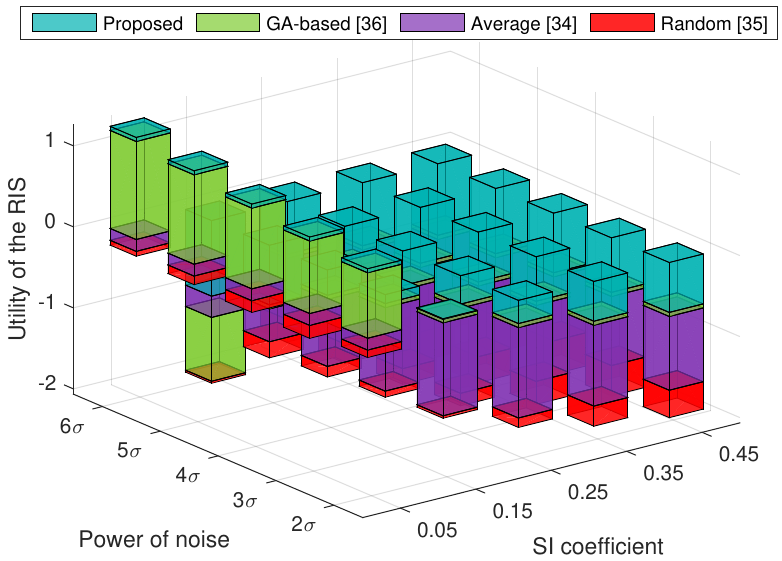}
}
\qquad
\hspace{-0.3in}
\subfigure[ ]
{
   \label{tuadd3}
    \includegraphics[width=0.6\columnwidth]{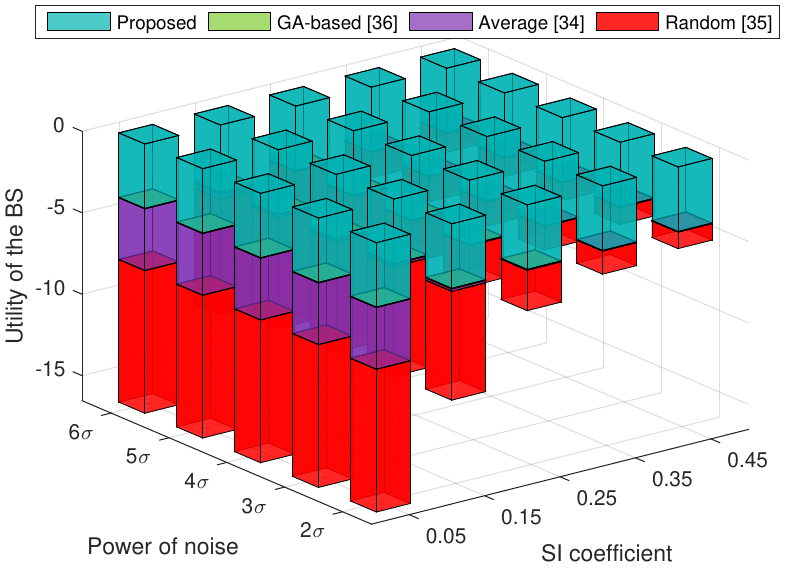}
}
\quad
\hspace{-0.3in}
\caption{Performance comparison of the proposed and GA-based optimization schemes. (a) The attacker's utility with different SI coefficients and attacking noise. (b) The RIS's utility with different SI coefficients and attacking noise. (c) The BS's utility with different SI coefficients and attacking noise.}
\vspace{-0.3cm}
\label{tuadd}
\end{figure*}

\textcolor{black}{Fig. \ref{tuadd} is the performance comparison of the proposed and GA-based optimization schemes. As shown in Fig. \ref{tuadd}, compared with the listed baselines, the proposed optimization can realize higher utilities for all the players within the established Stackelberg game. The reason is that GSSPI is a more efficient and targeted approach, fully exploiting the specific structure of the Stackelberg game and the players’ interaction. As a result, it converges faster and more reliably to a better solution than the GA (which explores the solution space more generally), average, and random allocation schemes. Moreover, as the attack intensity increases, the utilities of the attacker and the RIS both rise, while the base station’s utility remains stable. The reasons are that the attacker directly benefits from increased noise, enabling it to disrupt communication more. At the same time, the RIS can strategically adjust its power amplification gain to further mitigate interference and enhance the signal strength towards the legitimate receiver, while the BS can adjust its sensing data generation strategy to counteract the increased noise and maintain its performance.}

\subsubsection{Comparison of Stackelberg and Nash Equilibrium} 
\begin{figure*}[!htb]
\centering
\subfigure[ ]
{
    \label{tu13}
    \includegraphics[width=0.62\columnwidth]{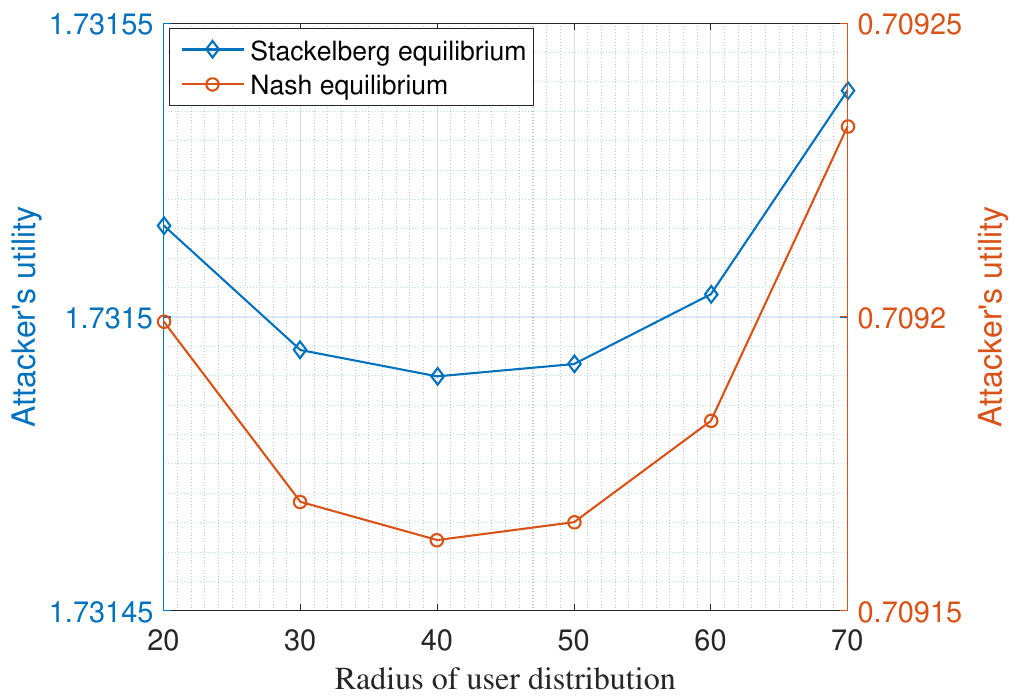}
}
\qquad
\hspace{-0.3in}
\subfigure[ ]
{
   \label{tu14}
    \includegraphics[width=0.62\columnwidth]{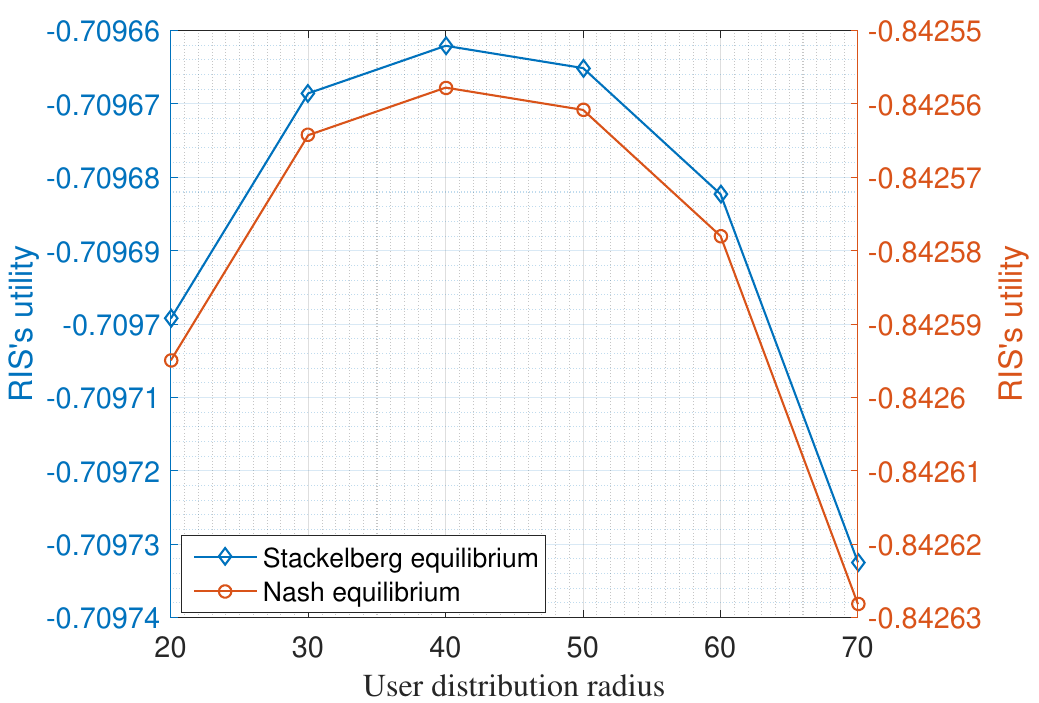}
}
\qquad
\hspace{-0.3in}
\subfigure[ ]
{
   \label{tu15}
    \includegraphics[width=0.62\columnwidth]{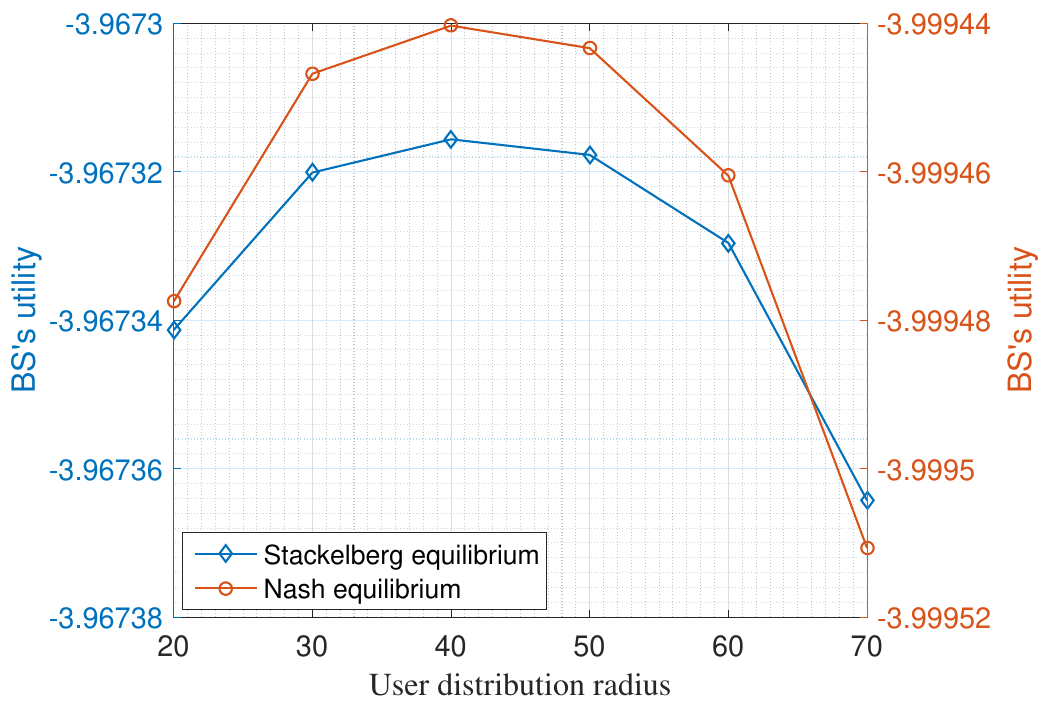}
}
\quad
\hspace{-0.3in}
\caption{Performance comparison of Stackelberg equilibrium and Nash equilibrium. (a) The attacker's utility with different equilibriums. (b) The RIS's utility with different equilibriums. (c) The BS's utility with different equilibriums. }
\label{tu131415}
\end{figure*}

Fig.~\ref{tu131415} is the performance comparison of Stackelberg equilibrium and Nash equilibrium. As can be seen from Fig. \ref{tu131415}, compared to Nash equilibrium, all the players can obtain the higher utilities with Stackelberg equilibrium. The reason is that the Stackelberg equilibrium considers the sequential causal relationship among all the players, reaching the better utility. In contrast, the Nash equilibrium considers only the static games between players \cite{yang2022aoi}, decreasing the utility in the focused scenario. Moreover, as the user-distribution radius expands, users closest to the BS enjoy higher SINR while those farther away suffer lower SINR. When the radius increases from 20 to 40 m, the average SINR rises, which lowers the attacker’s utility and raises the utilities of both the RIS and the BS. In contrast, extending the radius from 50 to 70 m reduces the average SINR, increasing the attacker’s utility and decreasing the utilities of the RIS and the BS.

\vspace{-0.2cm}
\section{Conclusion}\label{CC}
This paper aims at enhancing the ISAC performance of LAWNs for time-sensitive target under channel access attack. We have established the low-altitude network model under attack, deriving the communication data's SINR and sensing data's AoI expressions. Then, based on the defined utility functions and actions, we have formulated a Stackelberg game-based ISAC performance optimization problem, where the attacker acts as the leader, and legitimate drone with RIS and ISAC BS are regarded as the vice and first followers, respectively. With the proposed BI-based optimization algorithm, the Stackelberg equilibrium can be reached. The evaluation verifies the effectiveness and superiority of the proposed method, providing support for ensuring ISAC performance and building a reliable LAWNs. In the future, more constraints including energy consumption and control will be considered to further enhance LAWNs.
\vspace{-0.2cm}
\section*{Appendix A}
When both the generation and service of the sensing data follow exponential distributions, the M/M/1 queue model can be adopted to derive the AoI expression. Then, according to \cite{kaul2012real}, we have $\mathbb{E}\left[ B \right] = \frac{1}{\lambda }$ and $\mathbb{E}\left[ {{B^2}} \right] = \frac{1}{{{\lambda ^2}}}$. For the $i$-th generated sensing data, let $U_i$ and $O_i$ represent its waiting time and service time, its system time can be determined as:
\begin{equation}\label{1531564}
{T_i} = {U_i} + {O_i}.
\end{equation}

In addition, if the $(i-1)$-th sensing data completes service before the $i$-th sensing data is generated, $U_i =0$. Otherwise, if the $i$-th sensing data is generated while the $(i-1)$-th sensing data  is still queued or being served, $U_i = T_{i-1}-B_i$. Therefore, the waiting time of $i$-th sensing data is:
\begin{equation}\label{265666}
{U_i} = {\left( {{T_{i - 1}} - {B_i}} \right)^ + }.
\end{equation}

Based on \cite{miller1966probability}, for the M/M/1 queue model, when the system state is stable, the system times are stochastically identical, i.e., $T\mathop  = \limits^{st} {T_i}\mathop  = \limits^{st} {T_{i - 1}}$. The probability density function (PDF) of the system time is expressed as:
\begin{equation}\label{8484456256}
{f_T}\left( t \right) = {\gamma _{sense}}\left( {1 - \rho } \right){e^{ - {\gamma _{sense}}\left( {1 - \rho } \right)t}},\;\;t \ge 0.
\end{equation}

At this time, when $B_i=b$, the conditional expectation of the waiting time $U_i$ can be calculated as:

\begin{equation}\label{785625}
\begin{aligned}
\mathbb{E}\left[ {{U_i}|{B_i} = b} \right] &= \mathbb{E}\left[ {{{\left( {{T_{i - 1}} - {B_i}} \right)}^ + }|{B_i} = b} \right] \\
&= \mathbb{E}\left[ {{{\left( {T - b} \right)}^ + }} \right] \\
&= \int_b^\infty  {\left( {t - b} \right)} {f_T}\left( t \right)dt \\
&= \frac{{{e^{ - {\gamma _{sensing}}\left( {1 - \rho } \right)b}}}}{{{\gamma _{sensing}}\left( {1 - \rho } \right)}}.
\end{aligned}
\end{equation}

Since the service time $O_i$ is independent of $B_i$, we have:
\begin{equation}\label{956222}
\mathbb{E}\left[ {{T_i}{B_i}} \right] = \mathbb{E}\left[ {\left( {{U_i} + {O_i}} \right){B_i}} \right] = \mathbb{E}\left[ {{U_i}{B_i}} \right] + \mathbb{E}\left[ {{O_i}} \right]\mathbb{E}\left[ {{B_i}} \right].
\end{equation}

Note that $\mathbb{E}\left[ {{O_i}} \right] = \frac{1}{{{\gamma _{sense}}}}$, combined with (\ref{785625}), we have:
\begin{equation}\label{15465562125}
\mathbb{E}\left[ {{U_i}{B_i}} \right] = \int_0^\infty  {b\mathbb{E}\left[ {{U_i}|{B_i} = b} \right]} {f_{{B_i}}}\left( b \right)db = \frac{\rho }{{\gamma _{sense}^2\left( {1 - \rho } \right)}}.
\end{equation}

Finally, by integrating (\ref{956222}) and (\ref{15465562125}) to  (\ref{562}), we can obtain the average AoI of the sensing data as shown in (\ref{145123}). Hence, Theorem 1 is proved.

\section*{Appendix B}

When the sensing data is deterministically generated, and its service time is exponentially distributed, the sensing data AoI can be calculated under the D/M/1 queue model. Let $Cons$ represents the sensing data's fixed generation rate, we have $\lambda  = {1 \mathord{\left/
 {\vphantom {1 {Cons}}} \right.
 \kern-\nulldelimiterspace} {Cons}}$, $\mathbb{E}{{\left[ {{B^2}} \right]} \mathord{\left/
 {\vphantom {{\left[ {{B^2}} \right]} 2}} \right.
 \kern-\nulldelimiterspace} 2} = {{{{\left( {Cons} \right)}^2}} \mathord{\left/
 {\vphantom {{{{\left( {Cons} \right)}^2}} 2}} \right.
 \kern-\nulldelimiterspace} 2}$, and $\mathbb{E}\left[ {BT} \right] = Cons\mathbb{E}\left[ T \right]$. At this time, (\ref{562}) can be further expressed as:
\begin{equation}\label{15962569256}
AAoI = \frac{1}{{Cons}}\left[ {\frac{{{{\left( {Cons} \right)}^2}}}{2} + Cons\mathbb{E}\left[ T \right]} \right].
\end{equation}

In addition, according to \cite{kaul2012real}, the expectation of system can be calculated as:
\begin{equation}\label{14895521546}
\mathbb{E}\left[ T \right] = \mathbb{E}\left[ O \right] + \mathbb{E}\left[ U \right] = \frac{1}{{{\gamma _{sense}}}} + \frac{\delta }{{{\gamma _{sense}}\left( {1 - {\gamma _{sense}}} \right)}},
\end{equation}
where $\delta  \in \left[ {0,1} \right]$ can be determined by solving the equation $\delta  = {\mathcal{L}_B}\left( {{\gamma _{sense}}\left( {1 - \delta } \right)} \right)$, and $\mathcal{L}\left(  \cdot  \right)$ is the Laplace transform operator. Additionally, based on \cite{nelson2013probability}, for the deterministic sensing data generation, $\delta$ can further be expressed as:
\begin{equation}\label{695849514856}
\delta  = {e^{ - {\gamma _{sense}}\left( {1 - \delta } \right)Cons}} =  - \rho \Xi \left( { - \frac{{{e^{ - \frac{1}{\rho }}}}}{\rho }} \right),
\end{equation}
where $\Xi \left( \cdot \right)$ represents the Lambert-W function. Therefore, combined with (\ref{15962569256}) and (\ref{14895521546}), we can get (\ref{942563256}). Hence, Theorem 2 can be proved.

\section*{Appendix C}

Let the generation rate of the sensing data follow exponential distribution and the service time is deterministic, we can use M/D/1 queue model to derive the AoI expression of the sensing data. In this way, we have ${O_i} = Cons$. Since ${T_i} = {U_i} + Cons$, then:
\begin{equation}\label{158986589}
\mathbb{E}\left[ {{T_i}{B_i}} \right] = \mathbb{E}\left[ {{U_i}{B_i}} \right] + Cons\mathbb{E}\left[ {{B_i}} \right].
\end{equation}

Furthermore, similar to the calculation process of (\ref{785625}), at the condition $B_i=b$, the conditional expectation of the waiting time $U_i$ can be calculated as:
\begin{equation}\label{25962514856}
\begin{aligned}
\mathbb{E}\left[ {{U_i}|{B_i} = b} \right] &= \mathbb{E}\left[ {{{\left( {{T_{i - 1}} - {B_i}} \right)}^ + }|{B_i} = b} \right] \\
&= \mathbb{E}\left[ {{{\left( {T - b} \right)}^ + }} \right]  \\
&= \mathbb{E}\left[ {{{\left( {U + Cons - b} \right)}^ + }} \right] \\
&= \mathbb{E}\left[ U \right] + Cons - b - {\chi _{Cons}}\left( b \right){\xi _2}
\end{aligned}.
\end{equation}

Note that, based on \cite{yang2023stochastic}, for the M/D/1 queuing model, the expectation of waiting time $U$ can be expressed as:
\begin{equation}\label{51659658965489}
\mathbb{E}\left[ U \right] = \frac{{\rho Cons}}{{2\left( {1 - \rho } \right)}}.
\end{equation}

Finally, by integrating (\ref{15465562125}) and (\ref{25962514856}) into (\ref{562}), we can obtain (\ref{231223626}) , and hence Theorem 3 is proved.
\vspace{-0.2cm}

\bibliography{ref}{}
\bibliographystyle{IEEEtran}

\end{document}